\documentclass[12pt]{article}

\usepackage{textcomp}
\usepackage{chetnew}
\usepackage{tikz}
\usepackage{amssymb}

\newcommand{\CH}{\mathcal{H}}
\newcommand{\CD}{\mathcal{D}}
\newcommand{\CE}{\mathcal{E}}
\newcommand{\CQ}{\mathcal{Q}}
\newcommand{\CB}{\mathcal{B}}
\newcommand{\CC}{\mathcal{C}}
\newcommand{\CO}{\mathcal{O}}

\newcommand{\CI}{\mathcal{I}}
\newcommand{\CN}{\mathcal{N}}
\newcommand{\CS}{\mathcal{S}}

\preprint{RU-NHETC-2015-08, EFI-15-30, YITP-15-80}

\title{\vspace*{-10mm}Argyres-Douglas Theories, the Macdonald Index,\\[2mm] and an~RG~Inequality}

\author{Matthew Buican$^{\diamondsuit, 1,2}$ and Takahiro Nishinaka$^{\clubsuit, 1,3}$
}

\affiliation{$^1$NHETC and Department of Physics and Astronomy\\ Rutgers University, Piscataway, NJ 08854, USA \\ \smallskip$^2$Enrico Fermi Institute and Department of Physics, \\ The University of Chicago, Chicago, IL 60637, USA\\ \smallskip$^3$Yukawa Institute for Theoretical Physics\\ Kyoto University, Kyoto 606-8502, Japan \emails{$^{\diamondsuit}$mbuican@uchicago.edu, $^{\clubsuit}$takahiro.nishinaka@yukawa.kyoto-u.ac.jp}}

\abstract{We conjecture closed-form expressions for the Macdonald limits of the superconformal indices of the $(A_1, A_{2n-3})$ and $(A_1, D_{2n})$ Argyres-Douglas (AD) theories in terms of certain simple deformations of Macdonald polynomials. As checks of our conjectures, we demonstrate compatibility with two $S$-dualities, we show symmetry enhancement for special values of $n$, and we argue that our expressions encode a non-trivial set of renormalization group flows. Moreover, we demonstrate that, for certain values of $n$, our conjectures imply simple operator relations involving composites built out of the $SU(2)_R$ currents and flavor symmetry moment maps, and we find a consistent picture in which these relations give rise to certain null states in the corresponding chiral algebras. In addition, we show that the Hall-Littlewood limits of our indices are equivalent to the corresponding Higgs branch Hilbert series. We explain this fact by considering the $S^1$ reductions of our theories and showing that the equivalence follows from an inequality on monopole quantum numbers whose coefficients are fixed by data of the four-dimensional parent theories. Finally, we comment on the implications of our work for more general $\CN=2$ superconformal field theories.}

\date{September 2015}

\begin{document}
\setcounter{tocdepth}{2}

\maketitle
\toc

\newsec{Introduction}
When they were first constructed, Argyres-Douglas (AD) theories were defined as singular points on the Coulomb branches of certain $\CN=2$ gauge theories where mutually non-local BPS states become massless \rcite{Argyres:1995jj, Argyres:1995xn} (see also the generalizations in \rcite{Eguchi:1996vu, Gaiotto:2010jf}). This definition lead to remarkable insights into AD theories (e.g., the construction of the Coulomb branch chiral ring via the Seiberg-Witten curve and much more). On the other hand, the approach of starting from a UV gauge theory makes the computation of many observables in AD theories difficult. For example, since the superconformal $U(1)_R\subset U(1)_R\times SU(2)_R$ symmetry is emergent from this perspective, computing the superconformal index of an AD theory is highly non-trivial. At a more conceptual level, this construction obscures the inherent simplicity of AD theories by adding many extraneous degrees of freedom.

In a recent pair of papers \rcite{Buican:2015ina, Buican:2015hsa}, we advocated a different approach for computing the superconformal index of AD theories in the Schur limit.\foot{See also the very interesting orthogonal approach to the problem presented in \rcite{Cordova:2015nma}.} Our starting point was the class $\CS$ realization of AD theories as compactifications of the $(2,0)$ theory on a sphere, $\CC$, with an irregular singularity and at most one additional regular singularity \rcite{Bonelli:2011aa, Xie:2012hs} (see \rcite{Buican:2015ina} for a review). We then considered the topological quantum field theory living on $\CC$ (in this case, two-dimensional $q$-deformed $SU(2)$ Yang-Mills theory (YM) \rcite{Gadde:2011ik}) and defined a state in this theory corresponding to the irregular singularity (the states corresponding to regular singularities were already given in \rcite{Gadde:2011ik}). The resulting Schur index for the $(A_1, A_{2n-3})$ theory then has a simple representation as a sum over the components of the wave function of the irregular singularity in the basis of irreducible $SU(2)$ representations weighted by certain coefficients, while the index for the $(A_1, D_{2n})$ theory has a representation as a product of wave functions for the corresponding regular and irregular singularities \rcite{Buican:2015ina}.

An important aspect of our approach is that the irregular state is a simple and natural deformation of the regular state. As a result, it is possible to generalize our construction to limits of the index with more fugacities (and also, possibly, to the larger zoo of AD theories considered in \rcite{Xie:2012hs, Wang:2015mra}, although we leave the study of such theories to future work). 

In this note we propose just such a generalization to the Macdonald limit of the index. Recall that this limit is given by \rcite{Gadde:2011uv}\foot{In all formulas below, we follow the conventions of \rcite{Buican:2014qla}.}
\eqn{
\CI(q,t;\vec{x})={\rm Tr}_{\CH}(-1)^Fe^{-\beta\Delta}q^{2j_1}t^{R+r}\prod_i(x_i)^{f_i}={\rm Tr}_{\CH}(-1)^Fq^{2j_1}t^{R+r}\prod_i(x_i)^{f_i}~,
}[MacdonaldIndex]
where the trace is taken over the Hilbert space of local operators, $\CH$, $\Delta=\left\{\CQ_{2\dot-},(\CQ_{2\dot-})^{\dagger}\right\}$, $j_{1,2}$ are the $SO(4)$ spins, $R$ is the $SU(2)_R$ Cartan, $r$ is the $U(1)_R\subset U(1)_R\times SU(2)_R$ charge, and the $f_i$ are flavor charges. The fugacities $q$, $t$, and $x_i$ are complex numbers satisfying $|q|, |t|<1$ and $|x_i|=1$. We arrive at the last equality in \eqref{MacdonaldIndex} by well-known arguments which show that only operators satisfying $\Delta=E-2j_2-2R+r=0$ contribute to the index. Note that the Macdonald index counts precisely the same operators as the Schur index (therefore, we will interchangeably refer to the corresponding operators as constituting the Schur or Macdonald sector of the theory). However, it is more refined since it encodes two superconformal quantum numbers instead of one. Note that the Schur limit can be recovered by taking $t=q$. Similarly, the Hall-Littlewood (HL) limit can be reached by taking $q\to0$.\foot{The Macdonald and Schur limits count $1/4$ BPS operators. On the other hand, the HL limit counts only $3/8$ BPS operators (and in fact, as we will see below, it really only counts $1/2$ BPS operators in the $(A_1, A_{2n-3})$ and $(A_1, D_{2n})$ theories).}

Given this definition, we conjecture that the Macdonald indices of the $(A_1, A_{2n-3})$ and $(A_1, D_{2n})$ theories are given by\footnote{In this paper, we only consider the cases in which $n$ is a positive integer as in \cite{Buican:2015ina}.}
\eqn{
\mathcal{I}_{(A_1,A_{2n-3})}(q,t;x) = \sum_{\lambda=0}^\infty C_\lambda \, \tilde{f}_\lambda^{(n)}(q,t;x)~, \ \ \ \mathcal{I}_{(A_1,D_{2n})}(q,t;x,y) = \sum_{\lambda=0}^\infty \tilde{f}_\lambda^{(n)}(q,t;x)\, f_\lambda(q,t;y)~,
}[MainConjecture]
where $x$ and $y$ are fugacities for $U(1)$ and $SU(2)$ flavor subgroups, respectively (recall here that the flavor symmetries of the $(A_1,A_{2n-3})$ and $(A_1,D_{2n})$ theories are generically $U(1)$ and $SU(2)\times U(1)$, respectively). The coefficients, $C_{\lambda}$, in \eqref{MainConjecture} are given by
\begin{align}\label{Clambda}
C_\lambda &= \left[\frac{(t;q)_\infty}{(q;q)_\infty}\right]^{\frac{1}{2}}  \frac{P_\lambda(q,t;t^{\frac{1}{2}})}{(t^2;q)_\infty}~,
\end{align}
where $(x;q)_n \equiv \prod_{k=0}^{n-1}(1-q^kx)$, and $P_{\lambda}(q, t, x)$ is the normalized $A_1$ Macdonald polynomial
\begin{align}
P_\lambda(q,t;x) &= N_\lambda(q,t) \sum_{m=0}^{\lambda}\frac{(t;q)_m(t;q)_{\lambda-m}}{(q;q)_m(q;q)_{\lambda-m}}x^{2m-\lambda}~.
\label{eq:Macdonald}
\end{align}
Here the normalization factor is given by $N_\lambda(q,t) \equiv \sqrt{(1-t q^\lambda)\prod_{k=\lambda}^\infty (1-q^{k+1})(1-t^2q^k)}\times (q;q)_\lambda/(t;q)_\infty$ so that $\oint_{|x|=1} \frac{dx}{2\pi ix} \frac{(x^2;q)_\infty(x^{-2};q)_\infty}{2(tx^2;q)_\infty(tx^{-2};q)_\infty} P_\lambda(q,t;x)P_{\lambda'}(q,t;x) = \delta_{\lambda \lambda'}$. The factor $\tilde f^{(n)}_{\lambda}$ is the wave function of the irregular singularity of rank $n$, which we conjecture to be written as
\begin{align}\label{IrregWavefn}
\tilde{f}_\lambda^{(n)}(q,t;x) \equiv \left[\frac{(t;q)_\infty}{(q;q)_\infty}\right]^{\frac{1}{2}}P.E.\left[\frac{t}{1-q}\right]\tilde{P}_\lambda(q,t;x)~,
\end{align}
where  $P.E.\left[F(q,t;x_1,\cdots,x_\ell)\right] \equiv \exp\left(\sum_{k=1}^\infty \frac{F(q^k,t^k;x_1^k,\cdots,x_\ell^k)}{k}\right)$ for any function $F$, and the factor $\tilde{P}_\lambda(q,t;x)$ is the following deformation of the Macdonald polynomial:
\begin{align}
\tilde{P}_\lambda(q,t;x) &=  N_\lambda(q,t)\; t^{\frac{n\lambda}{2}}q^{\frac{n\lambda^2}{4}} \sum_{m=0}^{\lambda}\frac{(t;q)_m(t;q)_{\lambda-m}}{(q;q)_m(q;q)_{\lambda-m}}
q^{- n\left(\frac{\lambda}{2}-m\right)^2}x^{2m-\lambda}~.
\label{eq:def-Macdonald}
\end{align}
The wave function for the regular singularity, $f_{\lambda}$, takes the well-known form \rcite{Gadde:2011uv}
\begin{align}\label{RegSingWvFn}
f_\lambda(q,t;y) &= \left[\frac{(t;q)_\infty}{(q;q)_\infty}\right]^{\frac{1}{2}} P.E.\left[\frac{t}{1-q}\chi_{\text{adj}}^{su(2)}(y)\right]P_{\lambda}(q,t;y)~,
\end{align}
where the adjoint character of $su(2)$, $\chi_{adj}^{su(2)}(y)=y^2+1+y^{-2}$ (more generally, $\chi^{su(2)}_{\lambda}(y)=\sum_{m=0}^{\lambda}y^{2m-\lambda}$ for the spin $\lambda/ 2$ representation). Note that, as promised, the irregular wave function, \eqref{IrregWavefn}, is a simple deformation of the regular one, \eqref{RegSingWvFn}.

In our previous papers \rcite{Buican:2015ina, Buican:2015hsa}, we learned new things about AD theories by studying analytic properties of the Schur index. For example, we saw that the pole structure of the Schur limit---in particular the absence of certain poles---somewhat surprisingly encoded the spectrum of $\CN=2$ chiral primaries\foot{These are primaries that are annihilated by the full set of anti-chiral Poincar\'e supercharges (the $\bar\CE$ operators in the language of \rcite{Dolan:2002zh}; see also the earlier classification in \rcite{Dobrev:1985qv}). They are often referred to as \lq\lq Coulomb branch operators" since, in all known examples, their vevs parameterize the Coulomb branch of a theory.} even though these operators do not directly contribute in this limit!\foot{This surprise may be related to the fact that the authors of \rcite{Cordova:2015nma} were able to reproduce the Schur index by a BPS calculation on the Coulomb branch.} We took this fact as an indication of the simplicity of AD theories: the Schur sector does not consist of entirely new degrees of freedom but rather is highly constrained by the physics of the Coulomb branch. At the same time, we learned some lessons about more general $\CN=2$ SCFTs. For example, we saw that the $S^1$ reductions of theories with generic $\CN=2$ chiral ring spectra (i.e., theories with non-integer and non-half-integer dimensional $\CN=2$ chiral primaries) should have three-dimensional flavor symmetries (acting on $SU(2)_L\subset SU(2)_L\times SU(2)_R\simeq SO(4)_R$ charged primaries) that mix with the four-dimensional $U(1)_R$ symmetry upon compactification (at least as long as the four-dimensional theory has a Coulomb branch).

In a similar spirit, we will study the pole structure of the more refined Macdonald index below. We will see that away from the Schur limit (and also away from the HL limit, i.e., taking generic $t\ne q$) many of the poles that were missing (and whose absence encoded aspects of the Coulomb branch physics) reappear.\foot{It would therefore be interesting to understand if the Coulomb branch BPS physics arguments of \rcite{Cordova:2015nma} carry over to the Macdonald limit.} This behavior arises because of an intricate set of operator relations in the Schur/Macdonald sector that we will only scratch the surface of in this note (but which we will return to as part of a larger study \rcite{toappear}).

In addition, we extend our discussion of the dimensional reductions of the $(A_1, A_{2n-3})$ and $(A_1, D_{2n})$ theories that we initiated in \rcite{Buican:2015hsa}. However, instead of studying the resulting $S^3$ partition functions, we will instead examine aspects of the resulting three-dimensional indices. This study gives rise to a derivation of the equivalence of the HL limit of the $(A_1, A_{2n-3})$ and $(A_1, D_{2n})$ indices and the corresponding Hilbert series (for the $(A_1, A_{2n-3})$ theories, this series was computed in \rcite{DelZotto:2014kka}). Even more interestingly, we will find a sufficient condition for this equivalence that involves an inequality on the quantum numbers of three-dimensional monopole operators weighted by the mixing coefficients of the $U(1)_R$ symmetry descending from four dimensions with the topological symmetries of the dimensionally reduced theory. We will see that, as long as this mixing is sufficiently small, the HL limit of the index and the Hilbert series must agree. In generic $\CN=2$ theories, this result suggests a new criterion for the absence of index contributions due to exotic $\CD$ type HL operators (in the notation of \rcite{Dolan:2002zh}) and, possibly, a new constraint on RG flows between four and three dimensions that preserve eight supercharges.

Another important aspect of our previous work \rcite{Buican:2015ina} involved a comparison of the Schur indices of the $(A_1, A_3)$ and $(A_1, D_4)$ theories with the torus partition function of the corresponding two-dimensional chiral algebras (in the sense of \rcite{Beem:2013sza}). While the chiral algebras for the $(A_1, A_{2n-3})$ theories with $n>3$ and the $(A_1, D_{2n})$ theories with $n>2$ are still not known (see, however, \rcite{Cordova:2015nma} for the chiral algebras of many other AD theories), there are certain universal chiral sub-algebras that must be present in these theories on symmetry grounds. Moreover, we can use these chiral subalgebras to check the existence of certain operator equations predicted by our formulas. Note, however, that the chiral algebra only respects the superconformal quantum numbers of the Schur index and not all of those appearing in the Macdonald refinement. As a result, operator equations in the four-dimensional theory descend to null state relations in the two-dimensional chiral algebra that generally include terms whose four-dimensional pre-images violate the Macdonald quantum numbers (in this sense the Macdonald index also contains more refined information than the chiral algebra).\foot{A particularly trivial example of this phenomenon is given by those theories whose chiral algebras have a Sugawara construction. In that case, one finds an equation involving on one side a quadratic composite built out of currents (contributing at $\CO(t^2)$ to the Macdonald index) and involving on the other side a stress tensor (contributing at $\CO(qt)$ to the Macdonald index).}

The plan of this paper is as follows. In the next section we briefly motivate our conjectures for the Macdonald indices of the $(A_1, A_{2n-3})$ and $(A_1, D_{2n})$ theories. We then discuss certain checks applicable to the low-rank theories. In particular, we discuss checks arising from $S$-dualities involving these theories as building blocks \rcite{Buican:2014hfa, DelZotto:2015rca, Cecotti:2015hca} (see also the discussion in \rcite{Cecotti:2012va}).\foot{We hope that further generalizations of our work to the full class of theories considered in \rcite{Xie:2012hs, Wang:2015mra} will allow us to consider the non-self-dual $S$-duality discussed in \rcite{Buican:2014hfa} and its generalizations.} We then move on to more general checks involving the RG flow, the emergence of known Higgs branch relations, the equivalence of the HL limit and the Higgs branch Hilbert series, and the matching of operator relations to null states in the chiral algebras. In the following section we write our inequality on monopole quantum numbers and explain why the HL limits of the $(A_1, A_{2n-3})$ and $(A_1, D_{2n})$ indices should coincide with their Higgs branch Hilbert series. Finally, we include an initial discussion of the analytic properties of the index and the resulting consequences for the operator spectrum. We close with some conclusions.

\newsec{Motivating our Conjectures}
While the Macdonald index of a four-dimensional $\CN=2$ SCFT of class $\CS$ does not correspond to a correlator in a two-dimensional $q$-deformed YM theory on the punctured compactification curve, $\CC$, general arguments suggest that it should still correspond to a correlator  in {\it some} topological quantum field theory (TQFT) living on that curve \rcite{Gadde:2011uv}. Therefore, it is reasonable to assume that, just as in \rcite{Buican:2015ina}, we may find a state associated with the irregular singularities present in the class $\CS$ constructions of the $(A_1, A_{2n-3})$ and $(A_1, D_{2n})$ theories.

To give some further justification for our conjectures in \eqref{MainConjecture}, it is useful to first recall the Macdonald index of an $A_1$ theory corresponding to $\CC$ of genus $g$ with $m$ regular punctures \rcite{Gadde:2011uv}
\begin{align}
\mathcal{I}_{\mathcal{T}_{\mathcal{C}}}(q,t;x_1,\cdots,x_m)= \sum_{\lambda = 0}^\infty (C_\lambda)^{2-2g-m}\prod_{k=1}^{m}f_\lambda(q,t;x_k)~,
\end{align}
where the coefficients, $C_{\lambda}$, and the regular singularity wave function, $f_{\lambda}(q,t;x_k)$, are defined in \eqref{Clambda} and \eqref{RegSingWvFn} respectively. Note that the power of $C_{\lambda}$ is determined by the topology of $\CC$. Now, our formulas in \eqref{MainConjecture} come from considering one and two punctured spheres and replacing a regular singularity wave function with that of the irregular singularity. Therefore, in the case of the $(A_1, A_{2n-3})$ theory, we have $g=0$, $m=1$, and so it is natural for $C_{\lambda}$ to appear to the first power. On the other hand, in the case of the $(A_1, D_{2n})$ theory, we have $g=0$, $m=2$, and so it is natural for $C_{\lambda}$ to not appear in the corresponding index.

We can further justify the form of the irregular singularity wave function by noting
\eqn{
\lim_{t,q\to1}P.E.\left[-{t\over1-q}\right]\tilde f^{(n)}_{\lambda}(q,t;x)=\lim_{t,q\to1}P.E.\left[-{t\over1-q}\chi_{\rm adj}^{su(2)}(x)\right]f_{\lambda}(q,t;x)~.
}[Correspondence]
In other words, the irregular singularity wave function is a natural deformation of the regular singularity wave function and the corresponding Macdonald polynomials (up to a pre-factor that reflects the fact that the irregular singularity has $U(1)$ instead of $SU(2)$ flavor symmetry). More abstractly, we previously suggested \rcite{Buican:2015ina} that, in analogy with the description of irregular singularities as coherent states in the generalized AGT correspondence \rcite{Bonelli:2011aa, Gaiotto:2012sf, Kanno:2013vi}, the irregular singularity state appearing in the index should be thought of as an analog of a coherent state in the TQFT on $\CC$. Therefore, if we think of the limit $t, q\to1$ as a sort of classical limit, it is natural for the regular state wavefunction to coincide with the irregular state wavefunction up to an overall normalization.\foot{For instance, recall that in the basic example of the quantum mechanics of a simple harmonic oscillator, coherent states are the closest analogs to classical physics: the uncertainty, $\Delta x\Delta p$, is minimized in these states, and $\langle x\rangle$ and $\langle p\rangle$ are oscillatory with the classical frequencies and amplitudes.}

Another zeroth-order motivation for our conjecture is that it correctly reproduces the Schur limits of the $(A_1, A_{2n-3})$ and $(A_1, D_{2n})$ superconformal indices \rcite{Buican:2015ina, Cordova:2015nma}. Indeed, taking $t\to q$ in \eqref{MainConjecture}, we find that the irregular singularity wave function reduces to the expression in \rcite{Buican:2015ina}
\eqn{
\lim_{t\to q}\tilde f^{(n)}_{\lambda}(q,t;x)={1\over(q;q)_{\infty}}q^{nC_2(R_{\lambda})}{\rm Tr}_{R_{\lambda}}q^{-n(J_3)^2}a^{2J_3}~.
}[IrregWfnSchur]
In particular, the second and third factors in \eqref{IrregWfnSchur} are just the deformed Schur polynomials we studied in \rcite{Buican:2015ina}. Moreover, the $C_{\lambda}$ become the (normalized) coefficients of the irregular singularity wave function discussed in the expression for the $(A_1, A_{2n-3})$ Schur index \rcite{Buican:2015ina}
\eqn{
\lim_{t\to q}C_{\lambda}={\chi^{su(2)}_{\lambda}(q^{1\over2})\over(q^2; q)_{\infty}}=\CN(q)\left[{\rm dim} R_{\lambda}\right]_q~,
}[ClambdaSchur]
where $\left[{\rm dim} R_{\lambda}\right]_q=\chi^{su(2)}_{\lambda}(q^{1\over2})$ is the $q$-deformed dimension and $\mathcal{N}(q) \equiv 1/(q^2;q)_\infty$. The regular singularity wave function factor in the $(A_1, D_{2n})$ index, $f_{\lambda}$, behaves in the desired way under $t\to q$ by construction \rcite{Gadde:2011uv}.

As a final motivation, recall that one general property of the Macdonald index is that it is finite in the limit we take $q\to0$ with $t$ held fixed (this is the HL limit of the index).\foot{A priori, it need not be the case that the Macdonald index is finite if we instead send $t\to0$ and keep $q$ fixed. However, our conjectured forms of the $(A_1, A_{2n-3})$ and $(A_1, D_{2n})$ indices are also finite in this limit. In section \ref{analyticprop}, we will see what this statement implies for the operator spectra.} Indeed, this statement follows from the general expression \eqref{MacdonaldIndex} and the fact that all Macdonald operators have $j_1\ge0$. Note that this property is manifestly satisfied by our conjectures \eqref{MainConjecture} since $C_{\lambda}$, $\tilde f_{\lambda}^{(n)}$, and $f_{\lambda}$ are all finite in this limit.

\newsec{Low-Rank Checks}
In this section we perform checks of our conjecture \eqref{MainConjecture} that only apply to the subset of theories in our class that have rank zero or one.

\subsec{The $(A_1, A_1)$ and $(A_1, D_2)$ theories}

Let us first consider the $(A_1,A_1)$ and $(A_1,D_2)$ theories.
These are theories of free hypermultiplets (one in the case of the $(A_1, A_1)$ theory and two in the case of the $(A_1, D_2)$ theory). Relative to the generic $(A_1, A_{2n-3})$ and $(A_1, D_{2n})$ SCFTs we will consider below, these theories have enhanced flavor symmetry.

Since theories of free hypermultiplets have a Lagrangian description, we can evaluate their indices by direct computation. For the $(A_1, A_1)$ theory, we have
\begin{align}\label{A1A1index}
\CI_{(A_1, A_1)}(q,t;x)=\prod_{k=0}^\infty \frac{1}{(1-t^{\frac{1}{2}} q^k x)(1-t^{\frac{1}{2}}q^k x^{-1})}~,
\end{align}
where $x$ is a fugacity for the $Sp(1)\simeq SU(2)$ flavor symmetry.\footnote{We use the convention such that the character of a fundamental representation of $su(2)$ is $x + x^{-1}$.} On the other hand, our conjecture implies
\eqn{
\CI_{(A_1, A_1)}(q,t;x)=\sum_{\lambda=0}^{\infty}C_{\lambda}\tilde f^{(2)}_{\lambda}(q,t;x)~.
}[ConjA1A1]
We have checked that the two expressions \eqref{A1A1index} and \eqref{ConjA1A1} coincide to high perturbative order in $q$ and $t$. This agreement is highly non-trivial because our conjecture does not rely on a Lagrangian description of the theory. Note also that this agreement implies that the manifest $U(1)$ flavor symmetry in \eqref{ConjA1A1} is appropriately enhanced to $SU(2)$.

Next, let us consider the $(A_1, D_2)$ theory. Since this SCFT is a theory of hypermultiplets, its Macdonald index is similarly evaluated as 
\begin{align}\label{A1D2index}
\CI_{(A_1, D_2)}(q,t;x,y)=\prod_{s_1,s_2=\pm 1}\prod_{k=0}^\infty \frac{1}{(1-t^{\frac{1}{2}}q^k x^{s_1}y^{s_2})}~,
\end{align}
where $x$ and $y$ are fugacities for the $Sp(2)$ flavor symmetry.\footnote{We choose the basis of the Cartan subalgebra of $sp(2)$ so that $x$ and $y$ can be regarded as fugacities for $SU(2)\times SU(2)\subset Sp(2)$.} On the other hand, our conjecture implies that
\eqn{
\CI_{(A_1, D_2)}(q,t;x,y)=\sum_{\lambda=0}^{\infty}\tilde f^{(1)}_{\lambda}(q,t;x)f_{\lambda}(q,t;y)~.
}[ConjA1AD2]
Just as in the previous case, we have checked the equivalence of \eqref{A1D2index} and \eqref{ConjA1AD2} to high perturbative order in $q$ and $t$. Again, this agreement is highly non-trivial since the generic $(A_1, D_{2n})$ theory has $SU(2)\times U(1)$ flavor symmetry instead of $Sp(2)$.\footnote{As discussed below, the $(A_1, D_4)$ theory also has an enhanced flavor symmetry.}

\subsec{Symmetry Enhancement in the $(A_1, A_3)$ and $(A_1, D_4)$ theories}
\begin{table}
\begin{center}
\begin{tabular}{cl||cl}
\hline\hline
\hspace{1mm}monomial\hspace{2mm} & $su(2)$ representations & \hspace{1mm}monomial\hspace{2mm}& $su(2)$ representations \\ 
\hline
$1$ & ${\bf 1}$ & $q^2t$ & ${\bf 1},\,\; {\bf 3}$ \\
$t$ & ${\bf 3}$ & $t^4$ & ${\bf 9}$ \\
$t^2$ & ${\bf 5}$ & $qt^3$ & ${\bf 5},\,\; {\bf 7}$\\
$qt$ & ${\bf 1},\,\; {\bf 3}$ & $q^2t^2$ & ${\bf 1},\,\; 2\times {\bf 3},\,\; 2\times {\bf 5}$\\
$t^3$ & ${\bf 7}$ & $q^3t$ & ${\bf 1},\,\; {\bf 3}$\\
$qt^2$ & ${\bf 3},\,\;{\bf 5}$ & & \\
\hline\hline
\end{tabular}
\caption{The multiplicities of $su(2)$ representations at $\mathcal{O}(q^jt^k)$ for $0\leq j+k\leq 4$ in the Macdonald index of the $(A_1,A_3)$ theory.}
\label{table:multiplicity1}
\end{center}
\end{table}

\begin{table}
\begin{center}
\begin{tabular}{cl||cl}
\hline\hline
\hspace{1mm} monomial\hspace{2mm} & $su(3)$ representations &\hspace{1mm} monomial\hspace{2mm} & $su(3)$ representations \\ 
\hline
$1$ & ${\bf 1}$ &$q^2t$ & ${\bf 1},\,\; {\bf 8}$ \\
$t$ & ${\bf 8}$ & $t^4$ & ${\bf 125}$ \\
$t^2$ & ${\bf 27}$ & $qt^3$ & ${\bf 27},\,\;{\bf 35},\,\;\overline{\bf 35}$,\;\, {\bf 64}\\
$qt$ & ${\bf 1},\,\; {\bf 8}$ & $q^2t^2$ & ${\bf 1},\,\; 3\times {\bf 8},\,\;{\bf 10},\,\;\overline{\bf 10},\,\; 2\times {\bf 27}$\\
$t^3$ & ${\bf 64}$ & $q^3t$ & ${\bf 1},\,\;{\bf 8}$\\
$qt^2$ & ${\bf 8},\,\;{\bf 10},\,\;\overline{\bf 10},\,\; {\bf 27}$ & & \\
\hline\hline
\end{tabular}
\caption{The multiplicities of $su(3)$ representations at $\mathcal{O}(q^jt^k)$ for $0\leq j+k\leq 4$ in the Macdonald index of the $(A_1,D_4)$ theory.}
\label{table:multiplicity2}
\end{center}
\end{table}

Let us now consider the simplest interacting theories in our class of SCFTs: the $(A_1, A_3)$ and $(A_1, D_4)$ theories. One simple check of our conjectures in these cases is that we find the correct flavor symmetry enhancement to $SU(2)$ in the case of $(A_1, A_3)$ and $SU(3)$ in the case of $(A_1, D_4)$.\foot{These symmetries can be understood as a consequence of the corresponding flavor symmetries of the RG flows from the $SU(2)$ gauge theories with $N_f=2,3$ described in \rcite{Argyres:1995xn}.}

To see this enhancement, first recall that our conjecture implies
\eqn{
\CI_{(A_1, A_3)}(q,t;x)=\sum_{\lambda=0}^{\infty}C_{\lambda}\tilde f^{(3)}_{\lambda}(q,t;x)~, \ \ \ \CI_{(A_1, D_4)}(q,t;x,y)=\sum_{\lambda=0}^{\infty}\tilde f^{(2)}_{\lambda}(q,t;x)f_{\lambda}(q,t;y)~.
}[ConjA1A3]
The manifest flavor symmetries in these expressions are $U(1)$ for $(A_1,A_3)$ and $SU(2)\times U(1)$ for $(A_1,D_4)$. However, when we expand the above expressions in powers of $q$ and $t$, we see that the expansion coefficients are written in terms of characters of $su(2)$ in the case of $(A_1,A_3)$ and $su(3)$ in the case of $(A_1,D_4)$. We have checked this statement up to high perturbative order in $q$ and $t$. This result is in perfect agreement with the enhanced flavor symmetry of these theories described above. In tables \ref{table:multiplicity1} and \ref{table:multiplicity2}, we have written the low-dimensional operators and their representations under the flavor symmetry groups. In particular, note that the $\CO(t)$ contributions contain the expected moment maps transforming in the adjoint of $SU(2)$ and $SU(3)$ respectively (one can check that the only possible contributions at $\CO(t)$ arise from moment maps).

\subsec{The $(A_3, A_3)$ $S$-duality}
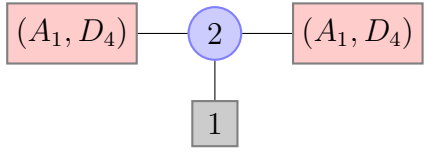
\begin{figure}
\begin{center}
\vskip .5cm
\begin{tikzpicture}[place/.style={circle,draw=blue!50,fill=blue!20,thick,inner sep=0pt,minimum size=7mm},transition/.style={rectangle,draw=black!50,fill=black!20,thick,inner sep=0pt,minimum size=6mm},transition2/.style={rectangle,draw=black!50,fill=red!20,thick,inner sep=0pt,minimum size=8mm},auto]

\node[transition2] (1) at (1.1,0) {\;$(A_1,D_4)$\;};
\node[place] (2) at (3,0) [shape=circle] {$2$} edge [-] node[auto]{} (1);
\node[transition2] (3) at (4.9,0)  {\;$(A_1,D_4)$\;} edge [-] node[auto]{} (2);
\node[transition] (9) at (3,-1.2) {$1$} edge[-] (2);
\end{tikzpicture}
\caption{The $(A_3,A_3)$ theory can be constructed by gauging a diagonal $SU(2)$ flavor symmetry of two $(A_1,D_4)$ theories and a fundamental hypermultiplet.}
\label{fig:quiver}
\end{center}
\end{figure}

In this subsection, we perform another check of the $(A_1, D_4)$ index. As discussed in \rcite{Buican:2014hfa} (see also the discussion in \rcite{Cecotti:2012va, DelZotto:2015rca}), we can consider taking two $(A_1, D_4)$ theories along with a doublet of hypermultiplets and gauging a diagonal $SU(2)$ flavor symmetry. The resulting coupling is exactly marginal, and the theory we obtain is identical to the $(A_3, A_3)$ theory \rcite{Buican:2014hfa}. This exactly marginal gauging implies that the Macdonald index of the $(A_3,A_3)$ theory can be written as
\begin{align}
\mathcal{I}_{(A_3,A_3)}(q,t;x,y,z) =& \oint\frac{dw}{2\pi i w}\Delta(w) \mathcal{I}_{\text{vect}}^{SU(2)}(q,t;w)\mathcal{I}_{\text{fund}}^{SU(2)}(q,t;x,w)
\nonumber\\
&\qquad \times \mathcal{I}_{(A_1,D_4)}(q,t;y,w)\, \mathcal{I}_{(A_1,D_4)}(q,t;z,w)~,
\label{eq:A3A3-index}
\end{align}
where $\Delta(w) = \frac{1}{2}(1-w^2)(1-w^{-2})$ is the measure factor and $\mathcal{I}_{\text{vect}}^{SU(2)}(q,t;w) = P.E.\left[\frac{-q-t}{1-q}\chi_{\text{adj}}^{su(2)}(w)\right]$ and $\mathcal{I}_{\text{fund}}^{SU(2)}(q,t;x,w) = P.E.\left[\frac{\sqrt{t}}{1-q}(x+x^{-1})\chi_{\text{fund}}^{su(2)}(w)\right]$ are the vector multiplet and hyper multiplet indices, respectively.

On the conformal manifold of the $(A_3,A_3)$ theory, there are various cusps where a dual gauge coupling goes to zero. These cusps are related to each other by the $S$-duality group, which acts on the three mass parameters (corresponding to the $U(1)^3$ flavor symmetry) via $S_3$ \rcite{Buican:2014hfa}. In terms of the corresponding flavor fugacities, this action gives rise to \rcite{Buican:2015ina}
\eqn{
x\to\sqrt{y\over z}~, \ \ \ y\to x\sqrt{yz}~, \ \ \ z\to{\sqrt{yz}\over x}~.
}[S3ActionA3A3]
Since the superconformal index is invariant under the $S$-duality, the index should satisfy the identity
\eqn{
\CI_{(A_3, A_3)}\left(q,t;x,y,z\right)=\CI_{(A_3, A_3)}\left(q,t;\sqrt{y/z},x\sqrt{yz},\sqrt{yz}/x\right)~.
}[IndexA3A3]
We have checked that our conjecture for $\mathcal{I}_{(A_1,D_4)}(q,t;x,y)$ correctly reproduces \eqref{IndexA3A3}, via \eqref{eq:A3A3-index}, up to a high perturbative order in $q$ and $t$. This result is highly non-trivial evidence for our conjecture.

\subsec{The $(A_2, A_5)$ $S$-duality}
Let us now perform a check of the $(A_1, D_6)$ index and (another check of) the $(A_1, A_3)$ index. Our test again involves an $S$-duality in a similar spirit to the one described in the previous subsection. To that end, recall that the authors of \rcite{DelZotto:2015rca} considered an SCFT built by taking an $(A_1, A_3)$ theory, an $(A_1, D_6)$ theory, and a fundamental hypermultiplet and gauging a diagonal $SU(2)$ flavor symmetry. This gauging turns out to be exactly marginal, and the resulting SCFT is the $(A_2, A_5)$ theory.\foot{This statement can be seen at the level of the hypersurface equation in \rcite{DelZotto:2015rca} or by a simple exercise in ($a$ and $c$) anomaly matching and flavor symmetry matching.} Like the $(A_3, A_3)$ theory discussed above, the conformal manifold of this theory has multiple cusps where the duality group ($SL(2, {\bf Z})$ in this case) acts on the two mass parameters of the theory (corresponding to the $U(1)^2$ flavor symmetry). Moreover, in \rcite{Cecotti:2015hca}, this action was argued to be via $S_3$.

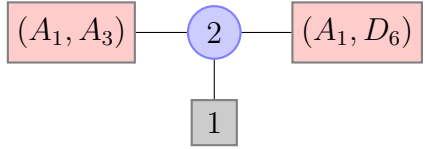
\begin{figure}
\begin{center}
\vskip .5cm
\begin{tikzpicture}[place/.style={circle,draw=blue!50,fill=blue!20,thick,inner sep=0pt,minimum size=7mm},transition/.style={rectangle,draw=black!50,fill=black!20,thick,inner sep=0pt,minimum size=6mm},transition2/.style={rectangle,draw=black!50,fill=red!20,thick,inner sep=0pt,minimum size=8mm},auto]

\node[transition2] (1) at (1.1,0) {\;$(A_1,A_3)$\;};
\node[place] (2) at (3,0) [shape=circle] {$2$} edge [-] node[auto]{} (1);
\node[transition2] (3) at (4.9,0)  {\;$(A_1,D_6)$\;} edge [-] node[auto]{} (2);
\node[transition] (9) at (3,-1.2) {$1$} edge[-] (2);
\end{tikzpicture}
\caption{The $(A_2,A_5)$ theory can be constructed by gauging a diagonal $SU(2)$ flavor symmetry of an $(A_1,A_3)$ sector, an $(A_1, D_6)$ SCFT, and a fundamental hypermultiplet.}
\label{fig:quiver}
\end{center}
\end{figure}

To see this symmetry at the level of the index, we first construct the $(A_2, A_5)$ index from the building blocks described in the previous paragraph
\begin{eqnarray}\label{A2A5}
\CI_{(A_2, A_5)}(q,t;x,y)=\oint{dw\over2\pi i w}\Delta(w)\CI^{SU(2)}_{\rm vect}(q,t;w)\ \CI^{SU(2)}_{\rm fund}(q,t;x,w)\nonumber\\ \times\ \CI_{(A_1, A_3)}(q,t;w^2)\ \CI_{(A_1, D_6)}(q,t; y,w)~.
\end{eqnarray}
It is then straightforward to check (as we have done perturbatively in $q$ and $t$) that the index is invariant under the following action on the flavor fugacities
\eqn{
x\to \frac{1}{\sqrt{x y}}~, \ \ \ y\to\sqrt{ x^3\over y}~,
}[S3actA2A5] 
i.e., that
\eqn{
\CI_{(A_2, A_5)}(q,t;x,y)=\CI_{(A_2, A_5)}(q,t;1/\sqrt{xy},\sqrt{x^3/y})~.
}[DualityA2A5]

Let us now find the action of $S_3$ more explicitly. Denote the transformation in \eqref{S3actA2A5} as $f$. The index is also symmetric under the transformation, $g$, which takes $x\to{1\over x}$ and leaves $y$ invariant. In terms of the corresponding chemical potentials, $m_{x,y}$, we have
\begin{eqnarray}\label{chempot}
f&:&\ \ \ m_x\to-{1\over2}(m_x+m_y)~, \ \ \ m_y\to{1\over2}(3m_x-m_y)~,\nonumber\\ g&:&\ \ \ m_x\to-m_x~, \ \ \ m_y\to m_y~.
\end{eqnarray}
Note that we have the relations
\eqn{
f^3=g^2=(fg)^2=1~,
}[S3defn]
which we recognize as the defining relations of $S_3$ (there cannot be additional relations since $S_3$ is the smallest non-Abelian group).

\newsec{General Checks}
In the following subsections, we perform checks of our conjectures that apply to all $(A_1, A_{2n-3})$ and $(A_1, D_{2n})$ theories.

\subsec{HL limit and Higgs branch relations}
\label{HLHiggs}
In this subsection, we recover the known Higgs branch relations for the $(A_1, A_{2n-3})$ and $(A_1, D_{2n})$ theories \rcite{Argyres:2012fu, DelZotto:2014kka, Buican:2015ina} from our index. The most efficient way to find these relations is to consider the HL limit of the index \rcite{Gadde:2011uv}: we keep $t$ fixed and set $q\to0$. In this limit, the index counts $3/8$ BPS operators and takes the vastly simplified form
\begin{align}\label{HLindex}
\text{Tr}_{\tilde{\mathcal{H}}}(-1)^F t^{R+r}\prod_{i}(x_i)^{f_i}~,
\end{align}
where the trace is over the subset of operators contributing to the Macdonald index with $j_1=0$.

A priori, there are two types of operators that can contribute in the HL limit. Using the nomenclature of \rcite{Dolan:2002zh}, the first type are highest $SU(2)_R$ weight primaries of $\hat\CB_{R}$ multiplets, and the second type are highest $SU(2)_R$ weight first-level superconformal descendants in the more exotic $\CD_{R(0,j_2)}$ multiplets. The $\hat\CB_{R}$ primaries (with highest $SU(2)_R$ weight) are the familiar $\CN=1$ chiral operators with scaling dimension $E=2R$ (they are anti-chiral with respect to the second set of Poincar\'e supercharges) whose vevs can parameterize the Higgs branch.\footnote{For example, the highest $SU(2)_R$ weight primaries of $\hat\CB_{1\over2}$ are free hypers and those of $\hat\CB_1$ are holomorphic moment maps for $\CN=2$ flavor symmetries.} On the other hand, the generic $\CD_{R(0,j_2)}$ multiplets are less familiar.\footnote{However, the multiplets with $R=j_2=0$ are well-known since they contain free vectors.}

In spite of the possibility of having exotic contributions to the HL index, we will find a consistent picture in which there are only $\hat\CB_R$ type contributions.\foot{The $D_{0(0,0)}$ multiplets cannot be present since they contain free vectors. Similarly, the arguments in \rcite{Buican:2014hfa} rule out the presence of $D_{0(0,j_2)}$ multiplets with $j_2>0$.} In particular, we will see that the HL limit of our conjecture \eqref{MainConjecture} coincides with the corresponding Higgs branch characters / Hilbert series (i.e., we will find a set of contributions equivalent to those coming from $\hat\CB_R$ operators modulo known Higgs branch constraints).  In section \ref{HLRsym}, we will argue that this agreement follows from general principles. Note also that this picture is consistent with the chiral algebras of the $(A_1, A_3)$ and $(A_1, D_4)$ theories described in \rcite{Buican:2015ina, Cordova:2015nma}. Indeed, from these chiral algebras, we know that there are no $\CD_{R(0,j_2)}$ multiplets in these theories. Finally, as we will see in section \ref{analyticprop}, simple analytic properties of our Macdonald indices rigorously forbid multiplets of the type $\bar\CD_{R(j_1,0)}$ with $R=0,{1\over2},1$ and $j_1\ge R$ (along with their conjugates).

\subsubsec{The $(A_1, A_{2n-3})$ theory}
To understand the above discussion more quantitatively, let us first take the HL limit of the $(A_1, A_{2n-3})$ index
\begin{align}\label{HLA1A2n3}
\mathcal{I}_{(A_1,A_{2n-3})}(0,t;x) = \sum_{\lambda=0}^\infty c_\lambda \tilde{f}_\lambda^{(n)}(0,t;x)~,
\end{align}
where $c_\lambda \equiv C_\lambda|_{q=0}$~. We see that $c_\lambda=\frac{1}{\sqrt{1-t}}t^{-\frac{\lambda}{2}}$ and $\tilde{f}_\lambda^{(n)}(0,t;x) = \frac{1}{\sqrt{1-t}}t^{\frac{n\lambda}{2}}(x^\lambda+x^{-\lambda})$ for $\lambda> 0$ while $c_0=\frac{1}{\sqrt{1-t^2}}$ and $\tilde{f}_0^{(n)}(0,t;x) = \sqrt{\frac{1+t}{1-t}}$. These formulas in turn imply that
\begin{align}
\mathcal{I}_{(A_1,A_{2n-3})}(0,t;x)
= \frac{1}{1-t}\left[1 + \sum_{\lambda=1}^\infty (x^\lambda + x^{-\lambda})t^{\frac{(n-1)\lambda}{2}}\right] 
= \frac{(1-t^{n-1})}{(1-t)(1-t^{\frac{n-1}{2}}x)(1-t^{\frac{n-1}{2}}x^{-1})}~.
\label{eq:A1An-HL}
\end{align}

To see the physical meaning of the above expression, recall that the Higgs branch operators of the theory are $M$ (with $R=1$ and $r=0$) and $N^{\pm}$ (with $R={n-1\over2}$ and $r=0$; the superscript is the $U(1)$ charge for $n>3$ and the $SU(2)$ flavor weight for $n=3$). These operators are subject to the Higgs branch constraint
\begin{align}
N^+ N^- = (-1)^{[\frac{n-1}{2}]}M^{n-1}~.
\label{eq:const1}
\end{align}
The quantum numbers of this constraint are $R=n-1$ and $r=0$. As a result, we see that \eqref{eq:A1An-HL} has a simple interpretation: it counts all products of $M$ and $N^{\pm}$ subject to \eqref{eq:const1}. Therefore the HL limit of our index agrees with the character of the Higgs branch chiral ring (which in turn agrees with the Higgs branch Hilbert series computation in \rcite{DelZotto:2014kka}). In section \ref{HLRsym} we will see that this agreement is no accident. This result also suggests that the Higgs branch chiral ring is equivalent to the HL chiral ring.\foot{For $(A_1, A_{2n-3})$ theories (with $n>3$), this statement holds modulo the possibility of miraculous cancelations between $\CD_{R(0,j_2)}$ multiplets of different spin.}

\subsubsec{The $(A_1, D_{2n})$ theory}
Let us now turn to the $(A_1, D_{2n})$ theory. Our conjecture \eqref{MainConjecture} implies that the HL index of the theory is given by
\begin{align}
\mathcal{I}_{(A_1,D_{2n})}(0,t;x,y) = \sum_{\lambda=0}^\infty \tilde{f}^{(n)}_{\lambda}(0,t;x)\,f_\lambda(0,t;y)~.
\end{align}
Note that $f_\lambda(0,t;y) = P^{HL}_{\lambda}(t;y)/\sqrt{1-t}(1-t y^2)(1-t y^{-2})$ where $P_\lambda^{HL}(t;y)$ is the HL polynomial: $P_\lambda^{HL}(t;y) = \chi^{su(2)}_\lambda(y) - t\chi^{su(2)}_{\lambda-2}(y)$ for $\lambda>0$ and $P_0^{HL}(t;x) = \sqrt{1+t}$. Using these expressions, we find that 
\begin{align}
\mathcal{I}_{(A_1,D_{2n})}(0,t;x,y)
= \frac{1}{(1-t)(1-ty^2)(1-ty^{-2})}\left[1 + t + \sum_{\lambda=1}^\infty t^{\frac{n\lambda}{2}}P^{HL}_\lambda(t;y)(x^\lambda + x^{-\lambda})\right].
\label{eq:HL-AD}
\end{align}

To understand this expression, recall that we have the following generators of the Higgs branch chiral ring: $M_i^{\ j}$ (with $R=1$ and transforming as $({\bf2}\otimes {\bf 2})_0$ under $SU(2)\times U(1)$), $L_i$ ($R=n/2$ and transforming as ${\bf2}_1$), and $\tilde L^i$ ($R=n/2$ and transforming as ${\bf2}_{-1}$). The relations among the generators are \cite{Argyres:2012fu, Buican:2015ina}
\begin{align}
\epsilon^{ik}\epsilon_{j\ell}M_i{}^j M_k{}^\ell = 0,\quad \epsilon^{ik}M_i{}^jL_k = 0,\quad \epsilon_{jk}M_i{}^j\tilde{L}^k = 0,\quad L_i\tilde{L}^j = (-1)^{[\frac{n}{2}]}M_i{}^j (M_k{}^k)^{n-1}~.
\label{eq:Higgs}
\end{align}
As in the case of the $(A_1, A_{2n-3})$ theory, it is now straightforward to show that \eqref{eq:HL-AD} agrees with the character of the Higgs branch chiral ring, ${\rm char}(\CH)$ (see appendix \ref{app:Higgs-A1Dn} for more detail). 
We will return to discuss this agreement from first principles in section \ref{HLRsym}. This result is also strong evidence that the Higgs branch chiral ring is equivalent to the HL chiral ring.\foot{As in the case of the $(A_1, A_{2n-3})$ theories (with $n>3$), this statement holds for $(A_1, D_{2n})$ theories with $n>2$ modulo the possibility of miraculous cancelations between $\CD_{R(0,j_2)}$ multiplets of different spin.}

\subsec{Operator equations beyond the Higgs branch and null states in the chiral algebra}
In this subsection, we would like to check certain operator constraints predicted by our conjectures that go beyond Higgs branch / HL chiral ring relations. In particular, we will focus on constraints involving the stress tensor multiplet (since the corresponding operators are universal) and / or derivatives of HL operators. Recall that the stress tensor multiplet is counted by the Macdonald index starting at $\CO(qt)$.

In theories with flavor symmetries (for simplicity, and in order to directly connect our discussion to the $(A_1, A_{2n-3})$ and $(A_1, D_{2n})$ SCFTs, we will assume the theories in question have either a simple global symmetry group or a single Abelian factor or both; generalizations of this ansatz are straightforward), the simplest operator constraints (that are counted by the Macdonald index) we can imagine involving the stress tensor multiplet and / or derivatives of HL operators take the form\foot{By \lq\lq simplest," we mean the lowest-dimensional independent operator constraints involving these operators and other universal operators.} 
\begin{eqnarray}\label{SimplestConstraint}
&&\kappa\cdot J_{+\dot+}^{11}M^{11I}+\kappa_f\cdot f^I_{\ JK}\left(M^{11J}\partial_{+\dot+}M^{11K}\right)+\kappa_1\cdot\partial_{+\dot+}M^{11}_{U(1)}M^{11I}+\kappa'_1\cdot M^{11}_{U(1)}\partial_{+\dot+}M^{11I}+\nonumber\\ &&+\kappa_f'\cdot d^I_{\ JK}\partial_{+\dot+}M^{11J}M^{11K}=0~,\nonumber \\ &&\tilde\kappa_1\cdot J_{+\dot+}^{11}M^{11}_{U(1)}+\tilde \kappa'_1\cdot\partial_{+\dot+}M^{11}_{U(1)}M^{11}_{U(1)}+\tilde\kappa_f\cdot\delta_{IJ}\partial_{+\dot+}M^{11I}M^{11J}=0~,\ \ \ \ \ \ \ \ \ \
\end{eqnarray}
where the $M^{11I}$ are holomorphic moment maps for the simple global symmetry group (the first two numbers in the superscript are $SU(2)_R$ indices set to highest weight, and the third is an adjoint flavor index), $M^{11}_{U(1)}$ is a holomorphic moment map for a $U(1)$ flavor symmetry, $J^{11}_{+\dot+}$ is the Schur component of the $SU(2)_R$ current (which sits in the stress tensor multiplet), the various $\kappa$'s are theory-dependent constants, the $f^I_{\ JK}$ are the structure constants of the flavor symmetry group, and $d^I_{\ JK}$ is the rank-three symmetric invariant tensor.\foot{In general, if the theory has additional $\hat\CC_{1(0,0)}$ multiplets, they may appear in \eqref{SimplestConstraint}.} There is no ordering ambiguity in \eqref{SimplestConstraint} since the $M^{11J}(x)M^{11K}(0)$, $M^{11}_{U(1)}(x)M^{11I}(0)$, $M^{11}_{U(1)}(x)M^{11}_{U(1)}(0)$, $J_{+\dot+}^{11}(x)M^{11}_{U(1)}(0)$, and $J_{+\dot+}^{11}(x)M^{11I}(0)$ OPEs do not contain singular terms.\foot{The absence of singularities in the OPEs of two holomorphic moment maps follows from a standard theorem regarding chiral ring operators. On the other hand, the absence of singularities in the $J_{+\dot+}^{11}(x)M^{11I}(0)$ and $J_{+\dot+}^{11}(x)M^{11}_{U(1)}(0)$ OPEs follows from $SU(2)_R$ symmetry and general Ward identities. Indeed, by $SU(2)_R$ conservation and dimensional considerations, the only potentially singular term in these OPEs would arise from a Schur operator in a multiplet of type $\hat\CB_2$. If such a term were present, then its chiral algebra image would contribute to the $\CO(z^{-1})$ term in the $T(z)J^I(0)$ (or $T(z)J_{U(1)}(0)$) chiral algebra OPEs. However, this term is fixed by Ward identities to be a descendant. Since the chiral algebra image of a $\hat\CB_2$ operator is a Virasoro primary \rcite{Beem:2013sza}, this is a contradiction.}

Before continuing, let us note that the free hypermultiplet theory (the $(A_1, A_1)$ theory) satisfies \eqref{SimplestConstraint}. Indeed, in this theory we have $J^{11}_{+\dot+}\sim \epsilon^{IJ}Q_I^1\partial_{+\dot+}Q_J^1$ and $M^{11I}\sim \sigma^I_{ij}Q^{1i}Q^{1j}$. Since the highest-weight component of the hypermultiplet is chiral, we can commute the hypermultiplets (and their derivatives). Therefore, it is trivial to see that these operators satisfy the first constraint in \eqref{SimplestConstraint} with $\kappa_1=\kappa_1'=0$ ($d^I_{\ JK}=0$ in this case).

Clearly we cannot use the HL index to study \eqref{SimplestConstraint}, since the operators in these relations do not contribute to this limit of the index. Moreover, in a general theory, it is non-trivial to use the Macdonald index to conclude that a constraint like \eqref{SimplestConstraint} is present. Indeed, while this constraint contributes to the Macdonald index negatively starting at $\CO(qt^2)$, there may instead be contributions of this same type coming from fermionic Macdonald operators. On the other hand, when we have a chiral algebra description of the Schur sector, we expect that \eqref{SimplestConstraint} will descend to a relation in this chiral algebra.

However, the null state equation corresponding to \eqref{SimplestConstraint}  in the chiral algebra will in general include operators whose four-dimensional pre-images have Macdonald quantum numbers that are different from the corresponding quantum numbers of the operators appearing in \eqref{SimplestConstraint}. The reason for this discrepancy is that the mapping of the four-dimensional Schur sector to the two-dimensional chiral algebra involves an $SU(2)_R$ twist \rcite{Beem:2013sza}.

To see this violation of Macdonald quantum numbers more explicitly, first recall that a Schur operator transforms as the highest $SU(2)_R$-weight component of an operator of $SU(2)_R$ spin $R$, $\CO^{i_1\cdots i_{2R}}$ (we suppress Lorentz and flavor indices for simplicity), i.e., $\CO_{\rm Schur}=\CO^{1\cdots1}$. However, the chiral algebra states are associated with non-trivial representatives of cohomology classes of a certain nilpotent supercharge that is a linear combination of a Poincar\'e supercharge and a special supercharge. As a result, in order for translations in the chiral algebra plane to be compatible with this cohomology structure, we must study the cohomology classes associated with \lq\lq twisted-translated" operators \rcite{Beem:2013sza}
\eqn{
\CO_A(z,\bar z)=u_{i_1}(\bar z)\cdots u_{i_{2R}}(\bar z)\CO_A^{i_1\cdots i_{2R}}(z,\bar z)~, \ \ \ u_{i_n}(\bar z)=(1, \bar z)~.
}[ImageSchur]
In particular, we see that, away from the origin, $\CO_A$ includes mixing with components of lower $SU(2)_R$ weight. Moreover, $\CO_A$ does not have definite dimension, $E$, or $SU(2)_R$ weight, $R$. It does, however, have definite $E-R$ (and definite flavor symmetry quantum numbers). As a result, $\CO_A$ can be associated with a definite Schur quantum number (this becomes the holomorphic dimension in the chiral algebra, $h=E-R$), but it does not have a full set of well-defined Macdonald quantum numbers. To find the representative chiral algebra operator, $\CO_A(z)$, we then work in the supercharge cohomology described above. The result of this process is written symbolically as $\CO_A(z)=\chi\left[\CO_A^{1\cdots1}\right]$, where $\chi\left[\cdots\right]$ is the map that takes a Schur operator in four dimensions and gives a chiral algebra operator in the associated two-dimensional theory.

The mixing in \eqref{ImageSchur} has additional implications for the images of composite operators. In particular,
\eqn{
\chi\left[\CO_1^{1\cdots1}\right]\chi\left[\CO_2^{1\cdots1}\right]=\chi\left[\CO_1^{1\cdots1}\CO_2^{1\cdots1}\right]+\sum_ka_{k}\cdot\chi\left[\CO_k^{1\cdots1}\right]~,
}[Ambiguity]
where the $a_k$ are certain theory-dependent constants determined by the particular $\CO^{1\cdots1}_1$ and $\CO^{1\cdots1}_2$ under consideration. In general, there are $a_k\ne0$, and so the operator product in four-dimensions does not translate directly into the operator product in the chiral algebra. These statements follow from considering the OPEs of the twisted-translated operators, $\CO_{1,2}$ defined as in \eqref{ImageSchur}, and noting that the normal ordered product (coming from the space-time independent term in the $\CO_1(z,\bar z)\CO_2(0)$ OPE) includes mixings with operators that appear in the OPE of lower $SU(2)_R$-weight components of $\CO_{1,2}^{i_1\cdots i_{2R}}$. Therefore, we should in general expect non-trivial mixing with chiral algebra images of $\CO_k^{1\cdots1}$ with differing Macdonald quantum numbers but the same Schur and flavor quantum numbers as the product $\CO_1^{1\cdots1}\CO_2^{1\cdots1}$. This fact makes finding the mapping between relations in four-dimensions and those in two dimensions highly non-trivial in general.

Let us now apply this discussion to the chiral algebra analog of \eqref{SimplestConstraint}. From the dictionary constructed in \rcite{Beem:2013sza}, we have the following 4d/2d maps
\eqn{
\chi\left[J_{+\dot+}^{11}\right]=-{1\over2\pi^2}T~, \ \ \ \chi\left[M^{11I}\right]={1\over2\sqrt{2}\pi^2}J^I ~,\ \ \  \chi\left[M^{11}_{U(1)}\right]={1\over2\sqrt{2}\pi^2}J_{U(1)}~,\ \ \ \chi\left[\partial_{+\dot+}\right]=\partial_z\equiv\partial~,
}[Univdictionary]
where $T$ is the two-dimensional holomorphic stress tensor, the $J^I$ (and $J_{U(1)}$) are currents of the two-dimensional Affine Kac-Moody algebra corresponding to the four-dimensional flavor symmetry, and $\partial$ is the holomorphic two-dimensional derivative (the normalization constants are determined in \rcite{Beem:2013sza}).

Given the above discussion, we see that the relations in \eqref{SimplestConstraint} descend to
\begin{eqnarray}\label{ChirAlgAnalog}
&-&{\kappa\over4\sqrt{2}\pi^4}\cdot TJ^I+{\kappa_f\over8\pi^4}\cdot f^I_{\ JK}\left(J^J\partial J^K\right)+{\kappa_1\over8\pi^4}\cdot\partial J_{U(1)}J^I+{\kappa_1'\over8\pi^4}\cdot J_{U(1)}\partial J^I+{\kappa_f'\over8\pi^4}d^I_{\ JK}\partial J^JJ^K+\nonumber \\ &+&\gamma\cdot\partial^2J^I=0~,\nonumber\\ &-&{\tilde\kappa_1\over4\sqrt{2}\pi^4}\cdot TJ+{\tilde\kappa_1'\over8\pi^4}\cdot\partial J_{U(1)}J_{U(1)}+{\tilde\kappa_f\over8\pi^4}\cdot \delta_{IJ}\partial J^IJ^J+\tilde\gamma\cdot\partial^2J_{U(1)}=0~.
\end{eqnarray}
In writing these equations, we have assumed that there are no higher-spin symmetries (i.e., that the theory is interacting). Indeed, if there are higher-spin symmetries, then we might find contributions of Schur operators in multiplets of type $\hat\CC_{0({1\over2},{1\over2})}$ in the $R=1$ OPEs of the $SU(2)_R$ currents and moment maps and in the $R=1$ OPEs of the moment maps with themselves.\foot{For example, in the case of the $(A_1, A_1)$ theory, such pollution contaminates the expression in \eqref{ChirAlgAnalog} (we have a Schur operator of the form $\partial_{+\dot+}Q^1_I\partial_{+\dot+}Q^1_J$ sitting in a $\hat C_{0({1\over2}, {1\over2})}$ multiplet).} The remaining operators that mix-in must be four-dimensional descendants and hence they must also be two-dimensional descendants (these are the operators multiplying $\gamma$ and $\tilde\gamma$ in \eqref{ChirAlgAnalog}).

Note that in the case of the $(A_1, A_{2n-3})$ and $(A_1, D_{2n})$ theories with $n>3$ and $n>2$ respectively, we do not know the full chiral algebras. However, we do know that there exist universal sub-algebras consisting of the affine Kac-Moody algebras corresponding to the $U(1)$ and $SU(2)\times U(1)$ flavor symmetries as well as, in the case of the $(A_1, A_{2n-3})$ theories, independent Virasoro sub-algebras (the $(A_1, D_{2n})$ theories all have Sugawara stress tensors).\foot{The fact that these are sub-algebras of the full chiral algebra follows from the fact that the corresponding singular OPEs are fixed by Ward identities.} Using these sub-algebras, we will argue that the null state in \eqref{ChirAlgAnalog} does not exist. We will see this picture is confirmed by our conjectures for the Macdonald indices. Therefore, we arrive at an internally consistent set of results.

\subsubsec{The $(A_1, A_3)$ and $(A_1, D_4)$ theories}
Let us first consider the $(A_1, A_3)$ SCFT. In this theory, we know that the only Macdonald / Schur operators consist of (derivatives of) products of the flavor symmetry moment maps and the $SU(2)_R$ current. Since there is a missing second ${\bf3}$ representation at $\CO(qt^2)$ in table \ref{table:multiplicity1}, we conclude that there must be a relation of the form \eqref{SimplestConstraint} (note that $d^I_{JK}=0$).

We can find the corresponding two-dimensional null relation as in \eqref{ChirAlgAnalog} by computing the following inner products for the states corresponding to the operators appearing in \eqref{ChirAlgAnalog}
\begin{eqnarray}\label{NullA1A3}
\langle TJ^I|TJ^J\rangle&=&k\left(4+{c\over2}\right)\delta^{IJ}=-{4\over3}\delta^{IJ}~,
\nonumber
\\\langle (J\partial J)^I|(J\partial J)^J\rangle&=& 8k\left(k+3\right)\delta^{IJ}=-{160\over9}\delta^{IJ}~, \nonumber
\\ \langle \partial^2J^I|\partial^2J^J\rangle&=&12k\delta^{IJ}=-16\delta^{IJ}~,
\\ \langle (J\partial J)^I|TJ^J\rangle&=&-4k\delta^{IJ}={16\over3}\delta^{IJ}~,\nonumber
\\ \langle \partial^2J^I|TJ^J\rangle&=&6k\delta^{IJ}=-8\delta^{IJ}~,\nonumber
\\ \langle\partial^2J^I|\left(J\partial J\right)^J\rangle&=&-16k\delta^{IJ}={64\over3}\delta^{IJ}~.\nonumber
\end{eqnarray}
In \eqref{NullA1A3}, $I, J=1,2,3$ is an $SU(2)$ adjoint index. We have defined $(J\partial J)^I\equiv f^{I}_{\ JK}\left(J^J\partial J^K\right)$, where $f^{I}_{\ JK}=i\sqrt{2}\epsilon^{I}_{\ JK}$. To arrive at the last set of equations in \eqref{NullA1A3}, we have used the fact that $k=-{4\over3}$ and $c=-6$ (see the discussion in \rcite{Buican:2015ina, Cordova:2015nma}). From these matrix elements it is straightforward to see that
\eqn{
TJ^I+{1\over2}\left(J\partial J\right)^I+{1\over6}\partial^2J^I=0~,
}[Relatqt3A3]
In particular, we find that in four-dimensions
\eqn{
J_{+\dot+}^{11}M^{I}-{1\over2\sqrt{2}}\left(M\partial_{+\dot+}M\right)^I=0~,
}[4dRelat3A3]
where we have defined $\left(M\partial_{+\dot+}M\right)^I\equiv f^{I}_{\ JK}\left(M^{J}\partial_{+\dot+}M^{K}\right)$. This result is an important additional check of our conjectured form of the Macdonald index.

We can proceed similarly for the $(A_1, D_4)$ theory. Note that there is a small subtlety: at $\CO(t^2)$ there is already a missing ${\bf8}$ representation due to the HL constraint \eqref{eq:Higgs} (with $n=2$). In particular, it is easy to check that the corresponding chiral algebra null vector is $d^A_{\ CD}J^CJ^D=0$ since $d^A_{\ CD}d^B_{\ EF}\langle J^CJ^D|J^EJ^F\rangle=10k\delta^{AB}\left({2\over3}k+1\right)=0$ (with $A, B=1,\cdots,8$ and $k=-3/2$).\foot{Mixing with $\partial J^A$ is forbidden by symmetry.} As a result, at level three, we have that $d^A_{\ CD}\partial J^CJ^D=0$. However, this is not an independent constraint since it follows from the level two constraint (and properties of the HL ring).

Now, checking table \ref{table:multiplicity2}, we see that there is an additional missing ${\bf 8}$ representation at $\CO(qt^2)$ and so we conclude there should be a relation of the form \eqref{SimplestConstraint}. We find the corresponding equation in the chiral algebra by computing the following matrix elements
\begin{eqnarray}\label{innerPA1D4}
\langle TJ^A|TJ^B\rangle&=&k\left(4+{c\over2}\right)\delta^{AB}=0~,
\nonumber\\\langle (J\partial J)^A|(J\partial J)^B\rangle&=& 6k\left(9+2k\right)\delta^{AB}=-54\delta^{AB}~,\nonumber
\\ \langle \partial^2J^A|\partial^2J^B\rangle&=&12k\delta^{AB}=-18\delta^{AB}~,\nonumber
\\ \langle (J\partial J)^A|TJ^B\rangle&=&-6k\delta^{AB}=9\delta^{AB}~,
\\ \langle \partial^2J^A|TJ^B\rangle&=& 6k\delta^{AB}=-9\delta^{AB}\nonumber~,
\\ \langle\partial^2J^A|\left(J\partial J\right)^B\rangle&=&-24k\delta^{AB}=36\delta^{AB}~,\nonumber
\end{eqnarray}
where $A, B=1,\cdots, 8$ are adjoint indices and we have used the fact that $k=-{3\over2}$ and $c=-8$ (see \rcite{Buican:2015ina, Cordova:2015nma}). We have normalized the structure constants in accord with the conventions of the previous example. As a result, we find the following null state
\eqn{
TJ^A+{1\over2}\left(J\partial J\right)^A+{1\over2}\partial^2J^A=0~.
}[OpEqnA1D4]
The corresponding four-dimensional operator equation is
\eqn{
J_{+\dot+}^{11}M^{11A}-{1\over2\sqrt{2}}\left(M\partial_{+\dot+}M\right)^A=0~.
}[4dRelat3D4]
This result is an important non-chiral check of our conjectured form of the $(A_1, D_4)$ index.

\subsubsection{Higher-rank theories}
\label{sec:HRk}
Let us first consider the $(A_1, A_{2n-3})$ theories with $n>3$. In this case, it is straightforward to expand our conjectured form of the index and observe that, subject to the assumptions that the only low-dimensional operators for generic $n$ are (derivatives of) products of the flavor moment maps and the $SU(2)_R$ current, there cannot be constraints of the type \eqref{SimplestConstraint} at $\CO(qt^2)$. In the chiral algebra we therefore expect there will not be a constraint of the form \eqref{ChirAlgAnalog} subject to the same assumptions. 

Indeed, we can compute the following matrix elements (from now on we change notation and take $J_{U(1)}\to J$ hoping that confusion will not arise)
\begin{eqnarray}\label{genericA2n3}
\langle TJ|TJ\rangle&=&k_1\left(4+{c\over2}\right)=-3n+11-{3\over n}~,\nonumber
\\ \langle\partial J\cdot J|\partial J\cdot J\rangle&=&8k_1^2=8~,\nonumber\\ \langle\partial^2J|\partial^2J\rangle&=&12k_1=12~, \\ \langle TJ|\partial^2J\rangle&=&6k_1=6~,\nonumber\\ \langle\partial J\cdot J|\partial^2J\rangle&=&\langle TJ|\partial J\cdot J\rangle= 0~,\nonumber
\end{eqnarray}
where, without loss of generality, we have set the $U(1)$ two point function, $k_1$, equal to unity, and we have used the fact that $c=-6n+14-{6\over n}$ \rcite{Buican:2015ina, Cordova:2015nma}. It is straightforward to check that the corresponding matrix is not degenerate, thus confirming our intuition from the Macdonald index.

Next, let us study the $(A_1, D_{2n})$ theories with $n>2$. These SCFTs have $SU(2)\times U(1)$ flavor symmetry. Expanding our conjectured form of the index subject to the assumptions we made in the $(A_1, A_{2n-3})$ case above, we see that there should not be constraints of the type \eqref{SimplestConstraint} at $\CO(qt^2)$ either in the $SU(2)$ adjoint channel or in the singlet channel. This calculation is straightforward but tedious and so we relegate it to appendix \ref{app:MatrixElt}.

We see that the constraint \eqref{SimplestConstraint} and the corresponding chiral algebra null state \eqref{ChirAlgAnalog} are very special. Unlike the singlet $\CO(t^2)$ Higgs branch constraints in \eqref{eq:const1} (for $n=3$), the first equation of \eqref{eq:Higgs}, and the Sugawara stress tensor equations in the corresponding chiral algebras, the constraints we have studied in this subsection apply to a finite number of theories. This relation is particularly intriguing because it only seems to apply to theories with low-dimensional Coulomb branch (complex dimension one or zero).

\subsec{The RG flow}
In this section, we will study the compatibility of our conjectures for the Macdonald indices with an intricate set of RG flows described in \rcite{Buican:2015ina}. Recall that these RG flows are triggered by giving a vev to some Higgs branch operator, $\CO$, of $SU(2)_R$ weight $R_{\CO}$ and charge $f_{k,\CO}$ under some $U(1)_k$ flavor symmetry (this generator may also be a Cartan of a non-Abelian flavor symmetry). Schematically, these RG flows are expressed as
\begin{align}
\mathcal{T}_{UV} \to \mathcal{T}_{IR} \oplus (A_1,A_1)~,
\end{align}
where $\mathcal{T}_{UV}$ is the UV SCFT while $\mathcal{T}_{IR}$ is the IR SCFT from which the decoupled axion-dilaton multiplet is excluded (for these RG flows the axion-dilaton is a free hypermultiplet, i.e., the $(A_1,A_1)$ theory).
Therefore, we will again use the prescription of \rcite{Gaiotto:2012xa} for relating the indices of the resulting IR endpoints of the RG flow to the indices of the corresponding UV endpoints
\eqn{
\CI_{\rm vect}^{-1}\cdot\CI_{IR}=-f_{i,\CO}\cdot {\rm Res}_{x_i=t^{-{R_{\CO}\over f_{\CO}}}\prod_{j\ne i}x_j^{-{f_{j,\CO}\over f_{i,\CO}}}}\left({1\over x_i}\CI_{UV}\right)~,
}[GRR]
where $\CI_{IR}$ is the index of the IR SCFT, and $\mathcal{I}_{\text{vect}}^{-1}\equiv P.E.\left[\frac{q+t}{1-q}\right]$ is the index of the decoupled axion-dilaton multiplet. Note that, since the indices we study here have an additional superconformal fugacity compared to the Schur indices we analyzed in \rcite{Buican:2015ina}, we will need some more powerful mathematical tools for isolating the residues in \eqref{GRR}. We will see that two particularly useful tools in our case are the $q$-binomial theorem and Bowman's generalization of Heine's transformation formula \rcite{Bowman}.

\subsubsec{Rewriting the indices}

Before beginning our analysis of the RG flows with endpoints in our class of theories, we would like to rewrite our conjectures for the Macdonald indices in such a way that we can easily extract the IR physics on the Higgs branch (i.e., so that we can straightforwardly apply \eqref{GRR} to an RG flow with $(A_1, A_{2n-3})$ or $(A_1,D_{2n})$ as the short-distance fixed point).

Let us start with the $(A_1,A_{2n-3})$ theory. Our conjecture \eqref{MainConjecture} implies that
\begin{align}
\mathcal{I}_{(A_1,A_{2n-3})}(q,t;x) = \frac{1}{(q;q)_\infty(t^2;q)_\infty}\sum_{\lambda=0}^\infty \tilde{P}_{\lambda}(q,t;x) P_\lambda(q,t;t^{\frac{1}{2}})~,
\label{eq:first-A1An}
\end{align}
where $P_\lambda$ and $\tilde{P}_\lambda$ are given in \eqref{eq:Macdonald} and \eqref{eq:def-Macdonald}. Since the $A_1$ Macdonald polynomial is the ultraspherical polynomial, there is a simple expression for $P_\lambda(q,t;t^{\frac{1}{2}})$: 
\begin{align}
P_\lambda(q,t;t^{\frac{1}{2}}) = N_\lambda(q,t)\frac{(t^2;q)_\lambda}{(q;q)_\lambda}t^{-\frac{\lambda}{2}}~,
\label{eq:x=t-to-half}
\end{align}
where $N_\lambda(q,t)$ is the normalization factor given below \eqref{eq:Macdonald}.
See appendix \ref{app:q-series} for a derivation of this expression. By combining the above two equations and using the infinite $q$-binomial theorem $\sum_{\lambda=0}^\infty \frac{(a;q)_\lambda}{(q;q)_\lambda}x^\lambda = \frac{(ax;q)_\infty}{(x;q)_\infty}$, we obtain
\begin{align}
\mathcal{I}_{(A_1,A_{2n-3})}(q,t;x) 
= \frac{1}{(t;q)_\infty^2}\sum_{m=0}^\infty \frac{(t;q)_m}{(q;q)_m} \left(x t^{\frac{n-1}{2}}\right)^m \left[\frac{(t^{\frac{n+1}{2}}q^{nm}x^{-1};q)_\infty}{(t^{\frac{n-1}{2}}q^{nm}x^{-1};q)_\infty} - q^{m}t \frac{(t^{\frac{n+1}{2}}q^{nm + 1}x^{-1};q)_\infty}{(t^{\frac{n-1}{2}}q^{nm+1}x^{-1};q)_\infty}\right]~.
\label{eq:resum1}
\end{align}
We will use this expression to evaluate \eqref{GRR} for the $(A_1,A_{2n-3})$ theory below.

Let us also rewrite our expression for $\mathcal{I}_{(A_1,D_{2n})}(q,t;x,y)$. Our conjecture \eqref{MainConjecture} implies that
\begin{align}
\mathcal{I}_{(A_1,D_{2n})}(q,t;x,y) = \frac{1}{(q;q)_\infty(t;q)_\infty (ty^2;q)_\infty(ty^{-2};q)_\infty}\sum_{\lambda=0}^\infty \tilde{P}_\lambda(q,t;x)P_\lambda(q,t;y)~.
\label{eq:first-A1Dn}
\end{align}
As in the previous case, we will rewrite this expression so that the residue computation in \eqref{GRR} can be easily performed. However, since the fugacity $y$ is now generic, we have a slightly more complicated expression than in the $(A_1,A_{2n-3})$ case. Indeed, as shown in appendix \ref{app:poles-A1Dn}, \eqref{eq:first-A1Dn} is rewritten as
\begin{align}
&\mathcal{I}_{(A_1,D_{2n})}(q,t;x,y) 
= \frac{(t^2;q)_\infty}{(t;q)_\infty^3(ty^2;q)_\infty(t/y^2;q)_\infty}\sum_{m_1,m_2=0}^\infty\!\!\! \frac{(t;q)_{m_1}(t;q)_{m_2}(t;q)_{|m_1-m_2|}(1-tq^{L_{m_1,m_2}})}{(q;q)_{\ell_{m_1,m_2}}(t^2;q)_{L_{m_1,m_2}}(q;q)_{|m_1-m_2|}}
\nonumber\\
&\quad\;\; \times A_{m_1,m_2}(q,t;x,y)\;\, {}_4\varphi_3 \!\left(
\begin{array}{cccc}
t & tq^{L_{m_1,m_2}+1} & q^{L_{m_1,m_2}+1} & tq^{|m_1-m_2|} \\
 & tq^{L_{m_1,m_2}} & t^2q^{L_{m_1,m_2}} & q^{|m_1-m_2|+1} \\
\end{array}
;\; q^{nm_1}t^{\frac{n}{2}}x^{-1}y
\right)~,
\label{eq:resum2}
\end{align}
where $L_{m_1,m_2} \equiv \text{max}(m_1,m_2),\, \ell_{m_1,m_2} \equiv \text{min}(m_1,m_2)$ and 
\begin{align}
A_{m_1,m_2}(q,t;x,y) \equiv
 q^{nm_1(L_{m_1,m_2}-m_1)}t^{\frac{nL_{m_1,m_2}}{2}}x^{2m_1-L_{m_1,m_2}}y^{-2m_2+L_{m_1,m_2}}~.
\end{align}
The function ${}_4\varphi_3$ is the basic hypergeometric series given in \eqref{eq:basic-hyper}.

\subsubsec{The $(A_1, A_{2n-3})\to(A_1, A_1)$ flow}
\label{subsec:A1An-residue}
Given our rewriting of the $(A_1, A_{2n-3})$ index in \eqref{eq:resum1}, we will study the following RG flow
\eqn{
(A_1, A_{2n-3})\to(A_1, A_1)~.
}[RGNm]
As discussed in \rcite{Buican:2015ina}, we can construct this flow by starting from the $(A_1, A_{2n-3})$ theory and turning on $\langle N^-\rangle\ne0$ with $\langle N^+\rangle=\langle M\rangle=0$ (recall that these operators were introduced around \eqref{eq:const1}). Since $N^-$ has charge $-1$ under the flavor $U(1)$ and has $SU(2)_R$ weight $R_{N^-}={n-1\over2}$, we see from \eqref{GRR} that we should compute the residue of $\mathcal{I}_{(A_1,A_{2n-3})}(q,t;x)$ at $x=t^{n-1\over2}$.

In \eqref{eq:resum1}, the residue at $x=t^{\frac{n-2}{2}}$ comes from the term proportional to
\begin{align}
\frac{(t^{\frac{n+1}{2}}x^{-1};q)_\infty}{(t^{\frac{n-1}{2}}x^{-1};q)_\infty}~.
\end{align}
All the other terms are finite at $x=t^{\frac{n-1}{2}}$, and moreover the sum of all such finite terms is convergent because of the conditions $|t|<1$ and $|q|<1$. Therefore, the residue is evaluated as
\begin{align}
Res_{x=t^{\frac{n-1}{2}}} \left[\frac{1}{x}\mathcal{I}_{(A_1,A_{2n-3})}(q,t;x)\right] = \frac{1}{(t;q)_\infty^2}Res_{x=t^{\frac{n-1}{2}}}\left[\frac{(t^{\frac{n+1}{2}}x^{-1};q)_\infty}{x(t^{\frac{n-1}{2}}x^{-1};q)_\infty}\right] = \left[\mathcal{I}_{vect}(q,t)\right]^{-1}~.
\end{align}
From \eqref{GRR}, we see this result is in perfect agreement with the RG flow in \eqref{RGNm} (recall that $\CI_{IR}=1$ in this case since the IR SCFT only consists of a decoupled axion-dilaton multiplet: the $(A_1, A_1)$ theory) and constitutes a strong check of our conjecture for the $(A_1, A_{2n-3})$  Macdonald index.

\subsubsection{The $(A_1,D_{2n})\to(A_1,A_1)\oplus (A_1,A_1)$ flow}
Let us now use \eqref{eq:resum2} to study the following RG flow
\eqn{
(A_1,D_{2n})\to(A_1,A_1)\oplus (A_1,A_1)~.
}[RGDA1A1]
Recall from the discussion in \rcite{Buican:2015ina} that we can generate this RG flow by turning on $\langle \tilde L^2\rangle\ne0$ and keeping the vevs of the remaining generators of the Higgs branch set to zero (these operators were discussed around \eqref{eq:Higgs}). Therefore, from \eqref{GRR}, we see that to study the flow \eqref{RGDA1A1}, we should take the residue of $\mathcal{I}_{(A_1,D_{2n})}(t,q;x,y)$ at $x=t^{\frac{n}{2}}y$.

To calculate this residue, we have to understand the analytic structure of the basic hypergeometric series. It turns out that there is a particularly useful rewriting of the basic hypergometric series due to Bowman \cite{Bowman}. Indeed, he found that
\begin{align}
{}_4\varphi_3\left(
\begin{array}{cccc}
a_0 & a_1 & a_2 & a_3\\
 & b_1 & b_2 & b_3\\
\end{array}
;\; z\right) = \frac{(a_0z;q)_\infty\prod_{k=1}^3(a_k;q)_\infty}{(z;q)_\infty\prod_{k=1}^3(b_k;q)_\infty}\sum_{\lambda=0}^\infty\frac{(z;q)_\lambda}{(a_0z;q)_\lambda(q;q)_\lambda}h_\lambda^{(3)}(\vec{b};\vec{a})~,
\end{align}
where
\begin{align}
h_{\lambda}^{(3)}(\vec{b};\vec{a}) \equiv \sum_{n_1+n_2+n_3=\lambda} \frac{(q;q)_\lambda}{(q;q)_{n_1}(q;q)_{n_2}(q;q)_{n_3}}\prod_{k=1}^3(a_k)^{n_k}(b_k/a_k;q)_{n_k}~.
\end{align}
This rewriting makes it manifest that, as a function of $z$, ${}_4\varphi_3(\vec{a};\vec{b};z)$ has simple poles at $q^kz = 1$ for $k=0,1,2,3,4,\cdots$. 

This property of the basic hypergeometric series implies that, in the sum over $m_1$ in \eqref{eq:resum2}, the terms with $m_1=0$ have simple poles at $x=t^{\frac{n}{2}}y$. All the other terms (with $m_1>0$) are finite at $x=t^{\frac{n}{2}}y$ (as long as the values of the other fugacities, $q$, $t$, and $y$, are generic). Moreover, the sum of all such finite terms is convergent due to the conditions $|q|<1,\,|t|<1$, and $|y|=1$. Therefore, the residue of $\mathcal{I}_{(A_1,D_{2n})}(q,t;x,y)$ at $x=t^{\frac{n}{2}}y$ is evaluated as
\begin{align}
Res_{x=t^{\frac{n}{2}}y}\left[\frac{1}{x}\mathcal{I}_{(A_1,D_{2n})}(q,t;x,y)\right] 
=&  \frac{1}{(q;q)_{\infty}(t;q)_\infty(ty^2;q)_\infty(t/y^2;q)_\infty}\sum_{m_2=0}^\infty \frac{(t;q)_{m_2}}{(q;q)_{m_2}} A_{0,m_2}(q,t;t^{\frac{n}{2}}y,y)
\nonumber\\
=& \left[\mathcal{I}_{\text{vect}}(q,t)\right]^{-1} \mathcal{I}_{(A_1,A_1)}(q,t;\mathfrak{y})~,
\end{align}
where $\mathfrak{y}\equiv \sqrt{t}y^2$ is the fugacity for the correct IR flavor symmetry as discussed in \cite{Buican:2015ina}.
This result is perfectly consistent with the RG flow described in \eqref{RGDA1A1}, since the IR SCFT with the axion-dilaton removed is just a free hypermultiplet, i.e., the $(A_1,A_1)$ theory.

\subsubsection{The $(A_1,D_{2n})\to (A_1,A_{2n-3})\oplus (A_1,A_1)$ flow}
Finally, we study the RG flow
\eqn{
(A_1,D_{2n})\to (A_1,A_{2n-3})\oplus (A_1,A_1)~.
}[RGDAnA1]
From the discussion in \rcite{Buican:2015ina}, we know that this RG flow can be initiated by turning on $\langle M_1^{\ 2}\rangle\ne0$ and keeping the vevs of the remaining Higgs branch generators zero. Using \eqref{GRR}, we see that this action corresponds to taking the residue of $\mathcal{I}_{(A_1,D_{2n})}(t,q;x,y)$ at $y=t^{\frac{1}{2}}$. In \eqref{eq:resum2}, only the prefactor $1/(ty^{-2};q)_\infty$ has a pole at $y=t^{\frac{n}{2}}$ (as long as the values of the other fugacities are all generic). This fact combined with \eqref{eq:first-A1Dn} imply that the residue of $\mathcal{I}_{(A_1,D_{2n})}(t,q;x,y)$ at $y=t^{\frac{1}{2}}$ is evaluated as:
\begin{align}
2Res_{y=t^{\frac{1}{2}}}\left[\frac{1}{y}\mathcal{I}_{(A_1,D_{2n})}(q,t;x,y)\right] =& \frac{2\left[\mathcal{I}_{\text{vect}}(q,t)\right]^{-1}}{(t^2;q)_\infty}Res_{y=t^{\frac{1}{2}}}\!\!\left[\frac{1}{y(ty^{-2};q)_\infty}\right]\sum_{\lambda=0}^\infty \tilde{P}_\lambda(q,t;x)P_\lambda(q,t;t^{\frac{1}{2}}) 
\nonumber\\
=& \left[\mathcal{I}_{\text{vect}}(q,t)\right]^{-1} \mathcal{I}_{(A_1,A_{2n-3})}(q,t;x)~.
\end{align}
This result is in perfect agreement with the RG flow in \eqref{RGDAnA1}.

\newsec{The HL Limit vs the Higgs Branch Hilbert Series and an RG Inequality}
\label{HLRsym}
In section \ref{HLHiggs}, we saw that the HL limits of the $(A_1, A_{2n-3})$ and $(A_1, D_{2n})$ indices agreed with the corresponding Higgs branch Hilbert Series. In this section, we would like to demonstrate that this agreement must occur.

Let us outline the argument. First recall that the index can be thought of as a twisted partition function on $S^1\times S^3$. We can then take the four-dimensional index and reduce it to the three-dimensional index via a ${\bf Z}_n$ ($n\to\infty$) quotient of the Hopf fiber of the $S^3$ \rcite{Benini:2011nc}. Since the HL index does not count operators with angular momentum quantum numbers along this fiber, we see that it is invariant under the reduction to three dimensions. Moreover, as we will see, the HL index reduces to the three-dimensional Higgs index. This latter index is then equivalent to the three-dimensional Higgs branch Hilbert series \rcite{Razamat:2014pta} (which is, in turn, equivalent to the four-dimensional Higgs branch Hilbert series).\foot{Strictly speaking, the above discussion holds only if the three-dimensional index does not blow up and only if we have identified the correct symmetries of the interacting IR fixed point (i.e., as long as the $S^1$ reduction is not \lq\lq bad"--- see the discussion in \rcite{Razamat:2014pta}). In our cases of interest, the dimensional reductions are all \lq\lq good."}

Note, however, that the above discussion is somewhat non-trivial in our case. Indeed, as observed in \rcite{Buican:2015hsa}, the $U(1)_R$ symmetries of the $(A_1, A_{2n-3})$ and $(A_1, D_{2n})$ theories flow to non-trivial linear combinations of the $SU(2)_L\subset SU(2)_L\times SU(2)_R\simeq SO(4)_R$ Cartans, $I_3^{L}$, and certain topological symmetries, $H_a^{\CC}$, of the $S^1$ reductions\foot{There is no such mixing only in the $(A_1, D_4)$ case.}
\eqn{
r\to I_3^{L}+c^aH_a^{\CC}~.
}[S1Creduction]
Therefore, we should check that this mixing does not spoil the above argument. Intuitively, we do not expect this to be the case since, as we saw in \rcite{Buican:2015hsa}, the mixing with topological symmetries was associated with the absence of certain poles involving non-HL operators. 

To understand the above discussion in more detail, first recall the form of the four-dimensional Lens space index (i.e., the partition function on $S^3/{\bf Z}_n\times S^1$)
\eqn{
\CI_{4d}={\rm Tr}_{S^3/{\bf Z}_n}(-1)^Fp^{j_2+j_1- r}q^{j_2-j_1- r}t^{ r+R}e^{-\beta(E-2j_2-2R+ r)}~,
}[IndexForm]
where the $n=1$ case is the usual superconformal index, while the limit $n\to\infty$ yields the three-dimensional index (here $\beta$ is the circumference of the $S^1$) \rcite{Benini:2011nc}. The quotient acts on the Hopf fiber via the phase, $\exp\left({2\pi i\over n}\right)$ (see \rcite{Razamat:2014pta} for a thorough review).

Let us now rewrite \eqref{IndexForm} using the substitution $p\to \sqrt{x\tilde x}y$, $q\to \sqrt{x\tilde x}y^{-1}$, and $t\to x$
\eqn{
\CI_{4d}={\rm Tr}_{S^3/{\bf Z}_n}(-1)^Fx^{j_2+R}\tilde x^{j_2-r}y^{2j_1}e^{-\beta(E-2j_2-2R+ r)}~.
}[IndexFormii]
Now, consider taking the HL limit. This amounts to taking $p,q\to0$ with $t$ fixed. In terms of our redefined fugacities, this is equivalent to $\tilde x\to0$ with $x$ fixed. Since the HL limit is independent of $j_1$, we drop the dependence on $y$. In particular, we find
\begin{eqnarray}\label{4d3dreductioniii}
\CI_{4d}^{\rm HL}&=&\lim_{\tilde x\to0}\CI_{4d}={\rm Tr}_{S^3/{\bf Z}_n}(-1)^Fx^{j_2+R}e^{-\beta(E-2j_2-2R+ r)}~,
\end{eqnarray}
and the final trace is over states with $j_2-r=0$.

We can now take the limit $n\to\infty$ in \eqref{4d3dreductioniii} and map the various charges appearing in \eqref{4d3dreductioniii} to three dimensions via the following dictionary
\eqn{
 r\to I_3^{L}+c^aH_a^{\CC}~,\ \ \ R\to I_3^{R}~, \ \ \ j_2\to j_2~.
}[4d3ddictionary]
Let us also define
\eqn{
\tilde E={1\over2}\left(E-I^L_3+c^aH_a^{\CC}\right)~.
}[DefntildeE]
Here $\tilde E$ is the three-dimensional scaling dimension for short multiplets that contribute to the (three-dimensional) index (while $E$ is the corresponding scaling dimension in four-dimensions).
We then find
\eqn{
\CI_{4d}^{HL}=\lim_{n\to\infty}\lim_{x\to\infty}\CI_{4d}={\rm Tr}_{S^2}x^{j_2+I_3^R}e^{-2\beta(\tilde E-I_3^R+I_3^L-j_2)}~.
}[HLim]
This is just the form of the Higgs index given in \rcite{Razamat:2014pta} (modulo mixings with topological symmetries, which were vanishing in the theories considered there).

To complete the argument, note that in the limit \eqref{HLim}, only operators satisfying $\tilde E=I_3^R+c^aH_a^{\CC}$ contribute to the index. In principle, we could imagine two types of operator contributions:  those from operators that are charged under the topological symmetries and those from operators that are not. Let us first consider the case of operators that are singlets under the topological symmetries. In this case, we need $\tilde E=I_3^R$. In our theories, such contributions come from three-dimensional Higgs branch scalars (since we have identified the correct symmetries of the IR theory).

Next, let us consider potential contributions from operators charged under the topological symmetries. A sufficient condition to rule out such contributions in our theories is to show that for any monopole primary, $\CO$, the following inequality holds
\eqn{
\tilde E(\CO)>(I_3^R+\sum_a c^aH_a^{\CC})(\CO)=\sum_a c^aH_a^{\CC}(\CO)~.
}[MonopoleEqn]
In \eqref{MonopoleEqn} we have used the fact that monopole primaries are $SU(2)_R$-neutral. In both the $(A_1, A_{2n-3})$ and $(A_1, D_{2n})$ cases, it is easy to check that the mirror of the above inequality holds for the matter operators in the mirror theories (see appendix \ref{app:Mineq}). Therefore, \eqref{MonopoleEqn} holds for all short multiplets of our $S^1$ reductions, and the Higgs branch Hilbert series and HL index agree as promised.

Note that \eqref{MonopoleEqn} is an inequality that depends on both ends of the $\beta\to0$ limit of the RG flow. Indeed, while the LHS and the $H_a^{\CC}(\CO)$ are determined by the three-dimensional long-distance physics, the mixing coefficients, $c^a$, are determined by the UV four-dimensional theory. It would be interesting to understand if \eqref{MonopoleEqn} is an inequality that holds for all RG flows from four dimensions to three dimensions that preserve eight supercharges or if it can be violated in some theories by sufficiently large $U(1)_R$ mixing with Coulomb branch symmetries.

\newsec{Comments on Analytic Properties of the Index}
\label{analyticprop}
In this section, we would like to make some preliminary comments on the analytic structure of the $(A_1, A_{2n-3})$ and $(A_1, D_{2n})$ Macdonald indices. It turns out that these indices have the following properties:

\begin{itemize}
\item[{\bf(i)}] The coefficients of all the terms in the $q$, $t$ expansion are {\it positive} integers.

\item[{\bf(ii)}] The $t\to0$ with $q$ fixed limits of our indices are finite and equal to unity.

\item[{\bf(iii)}] Certain poles associated with non-HL operators that were absent in the Schur limit reappear in the Macdonald limit.
\end{itemize}

\noindent

Let us first examine {\bf (i)}. For the $(A_1,A_{2n-3})$ theories, it follows from \eqref{eq:def-Macdonald}, \eqref{eq:first-A1An}, and \eqref{eq:x=t-to-half} that 
\begin{align}\label{eq:sign-A1An}
\mathcal{I}_{(A_1,A_{2n-3})}(q,t;x) 
= \sum_{\lambda=0}^\infty \sum_{m=0}^\lambda \frac{(1-q^\lambda t)}{(q^m t;q)_\infty(q^{\lambda-m}t;q)_\infty}\frac{1}{(q;q)_m(q;q)_{\lambda-m}}q^{nm(\lambda-m)}t^{\frac{(n-1)\lambda}{2}}x^{2m-\lambda}~.
\end{align}
It is now clear that, when expanded in powers of $q$ and $t$, the index has only positive integer coefficients. It is straightforward to prove the same statement for the $(A_1,D_{2n})$ theories. Indeed, as shown in appendix \ref{app:q-series}, the $(A_1,D_{2n})$ index can be rewritten as
\begin{align}
\mathcal{I}_{(A_1,D_{2n})}(q,t;x,y) 
= &\frac{1}{(t;q)_\infty}\sum_{\lambda=0}^\infty \frac{t^{\frac{n\lambda}{2}}}{(q;q)_\lambda}\sum_{m_1,m_2=0}^\lambda \left[\!\!
\begin{array}{c}
\lambda\\
m_1\\
\end{array}
\!\!\right]_q \left[\!\!
\begin{array}{c}
\lambda\\
m_2
\end{array}
\!\!\right]_q \frac{(1-q^\lambda t)}{(q^{m_1}t;q)_{\infty}(q^{\lambda-m_1}t;q)_\infty}
\nonumber\\[1mm]
&\qquad \qquad \times   \sum_{k=0}^\infty \frac{(q^{\lambda-m_2}ty^2)^k}{(q;q)_k(q^{m_2+k}ty^{-2};q)_\infty} q^{nm_1(\lambda-m_1)} x^{2m_1-\lambda} y^{\lambda-2m_2}~,
\label{eq:sign-A1Dn}
\end{align}
where $\scriptsize{\left[\!\!
\begin{array}{c}
\lambda\\
m
\end{array}
\!\!\right]_q} \equiv (q;q)_\lambda/\{(q;q)_{m}(q;q)_{\lambda-m}\}$ is the $q$-binomial coefficient and therefore a polynomial in $q$ with positive integer coefficients. It is manifest in \eqref{eq:sign-A1Dn} that, in the expansion in powers of $q$ and $t$, all the coefficients are positive integers.

These statements are consistent with the conjecture that all the Schur operators in the $(A_1,A_{2n-3})$ and $(A_1,D_{2n})$ theories are bosonic operators. In particular, they are consistent with the conjecture that the only Schur generators in our theories are the $SU(2)_R$ current, the flavor currents, and the baryons.\foot{This statement is therefore also consistent with a conjecture about the $(A_1, A_{2n-3})$ chiral algebra with $n>2$ mentioned to us by L.~Rastelli.} More generally, it is consistent with all the checks we performed above (we were not forced to include fermionic degrees of freedom). 

Let us now discuss property {\bf(ii)}. It is straightforward to check that \eqref{eq:sign-A1An} and \eqref{eq:sign-A1Dn} imply
\eqn{
\lim_{t\to0}\mathcal{I}_{(A_1,A_{2n-3})}(q,t;x)=\lim_{t\to0}\mathcal{I}_{(A_1,D_{2n})}(q,t;x) =1~,
}[t0lim]
where we have held $q$ fixed. From this simple fact, it immediately follows that:
\begin{itemize}
\item[$\bullet$] The $(A_1, A_{2n-3})$ and $(A_1, D_{2n})$ theories do not have multiplets of type $\bar\CD_{R(j_1,0)}$ with $R=0,{1\over2},1$ and $j_1\ge R$ or conjugate multiplets of type $\CD_{R(0,j_2)}$ with $R=0,{1\over2},1$ and $j_2\ge R$.
\end{itemize}
To understand this statement, first note that the $\bar\CD_{R(j_1,0)}$ multiplets with $R=0,{1\over2},1$ (except for $\bar\CD_{1(0,0)}$) and their conjugate multiplets do not participate in recombination rules that give long multiplets \rcite{Dolan:2002zh} (here we are using the absence of higher-spin symmetries). Their single letter contributions to the index are
\eqn{
\CI_{\bar D_{R(j_1,0)}}=(-1)^{2j_1+1}{t^{R-j_1}q^{2j_1+1}\over1-q}~.
}[Dcont]
Note that \eqref{Dcont} is singular in the $t\to0$ limit of the index for $j_1>R$. Moreover, it is easy to check that these are the only non-recombinant Macdonald multiplets that contribute singularly in the limit $t\to0$ and that these multiplets do not experience \lq\lq accidental" cancelations when $q$ is allowed to remain arbitrary. Since the limit \eqref{t0lim} is non-singular, it is then impossible to have $\bar\CD_{R(j_1,0)}$ multiplets with $R=0,{1\over2},1$ and $j_1> R$ (similar statements hold for the corresponding conjugate $\CD$ multiplets). To rule out the remaining multiplets with $R=j_1=0,{1\over2},1$, we note that these multiplets would contribute a $q$-dependent piece to \eqref{t0lim}. The absence of higher-spin symmetries forbid any canceling contributions to the lowest-dimensional such contributions. Therefore, these multiplets are also absent.

These results are entirely consistent with the picture we have described so far. Moreover, the above conclusions extend and confirm the results of \rcite{Buican:2014qla}, which imply the absence of multiplets of type $\CD_{0(0,j_2)}$ and $\bar\CD_{0(j_1,0)}$ in our theories.

Let us now turn to the property {\bf (iii)}. To see an example of {\bf(iii)}, consider the index of the $(A_1, A_{2n-3})$ theory for $n\geq 3$. Let us study the residue at $x=q^2t^{\frac{n-1}{2}}$ associated with some non-HL operator, $\CO$. From the expression \eqref{eq:resum1}, we see that the residue of $\mathcal{I}_{(A_1,A_{2n-3})}(q,t;x)$ is evaluated as
\begin{align}
Res_{x=q^2t^{\frac{n-1}{2}}}\left[\frac{1}{x}\mathcal{I}_{(A_1,A_{2n-3})}(q,t;x)\right]
= -\left[\mathcal{I}_{\text{vect}}(q,t)\right]^{-1} \frac{(t-q)(1-t)q^2}{(1-q)(1-q^2)}~.
\end{align}
Note that this residue is indeed zero for $t\to q$.

The residue can vanish in this limit because, while some constraint in the theory has the same Schur quantum numbers as $\CO$, it has different Macdonald quantum numbers. One way in which such a constraint may arise is if we have a relation of the form
\eqn{
(J_{+\dot+}^{11})^k(N^-)^k+\cdots=0~,
}[Tbaryon]
where $J_{+\dot+}^{11}$ is the Schur component of the $SU(2)_R$ current and $\CO$ is a second derivative of an HL operator. Indeed, we saw an example of such a constraint in \eqref{4dRelat3A3} for $n=3$ (and $k=1$). The basic point is that although $J_{+\dot+}^{11}$ has the same Schur quantum numbers as two derivatives, it has different Macdonald quantum numbers. As a result, constraints like \eqref{Tbaryon} will not cancel a pole associated with the second derivative of an HL operator in the generic Macdonald limit but may cancel that pole in the Schur limit.

Constraints of the form \eqref{Tbaryon} are intriguing. Indeed, the $SU(2)_R$ current knows about all sectors of the AD theory, since any multiplet has operators charged under $SU(2)_R$ (in the case of \lq\lq Coulomb branch" operator multiplets, these are superconformal descendants). Moreover, in our previous work \rcite{Buican:2015hsa} we saw that the absent poles in the Schur limit were intimately connected with the fact that the quantum numbers of the \lq\lq Coulomb branch" operators were secretly encoded in the index. Therefore, understanding the physics associated with constraints of the form \eqref{Tbaryon} may point the way to constructing the full index of our AD theories.

\newsec{Discussion}
We have generalized our construction of the Schur indices of the $(A_1, A_{2n-3})$ and $(A_1, D_{2n})$ theories to the Macdonald limit. In performing various checks of our conjectures, we arrived at an intriguing inequality on monopole quantum numbers in the dimensional reductions, and we found some interesting operator relations involving the $SU(2)_R$ current and various HL operators. We expect both these results to be useful in studying generic $\CN=2$ theories. A natural (partial) list of future directions include:

\begin{itemize}
\item[$\bullet$] Generalize our formulas to include the final superconformal fugacity. Better understanding the operator relations involving the $SU(2)_R$ current and the HL operators might be useful.

\item[$\bullet$] Find the full set of theories which satisfy \eqref{SimplestConstraint}. Do the corresponding chiral algebras necessarily have stress tensors given by the Sugawara construction? Is the list of these theories finite? Perhaps further understanding this equation can give additional insight into the possible set of low-rank $\CN=2$ theories with flavor symmetries (recall that this equation only held in the rank zero and one theories we studied).\foot{Note that the $(A_1, A_2)$ theory rather trivially satisfies \eqref{SimplestConstraint} since it has no flavor symmetries (we can think of such theories as having vanishing moment maps). See \rcite{Liendo:2015ofa} for an interesting recent discussion of this SCFT. However, if we are willing to accept such trivial solutions, then clearly we can find an infinite number of solutions (e.g., all the $(A_1, A_{2n})$ theories).}

\item[$\bullet$] Can the inequality we found on monopole operator quantum numbers, \eqref{MonopoleEqn},  be violated in more general theories? Are there interesting theories for which this is an equality? In this case, \eqref{MonopoleEqn} might be an interesting generalization of the topological criterion for the appearance of $\CD$ multiplets in the special subset of class $\CS$ theories with only regular singularities \rcite{Gadde:2011uv}.

\item[$\bullet$] On the other hand, if \eqref{MonopoleEqn} cannot be violated, then it may imply interesting constraints on UV versus IR physics of general RG flows from 4d to 3d preserving eight supercharges. In this sense it would be somewhat similar in spirit to other constraints on the RG flow that are already known (e.g., \rcite{Zamolodchikov:1986gt,Komargodski:2011vj,Komargodski:2011xv}) or conjectured (e.g., \rcite{Intriligator:2005if, Buican:2011ty,Buican:2012ec}). The main novelty would be a constraint on flows between dimensions. See appendix \ref{app:Mineq} for a very non-trivial check of this inequality in our theories (especially in the large central charge limit).

\item Further study the mathematical meaning of our deformation of the Macdonald polynomials given in \eqref{eq:def-Macdonald}. Since the $A_1$ Macdonald polynomials are equivalent to ultraspherical polynomials, our deformation can also be regarded as a deformation of these latter polynomials.
\end{itemize}

\bigskip\bigskip\bigskip
\ack{ \bigskip
We are grateful to C.~Beem, L.~Rastelli, and S.~Razamat for interesting discussions and communications. M.~B. would like to thank the Aspen Center for Physics, which is supported by National Science Foundation grant PHY-1066293, for a truly wonderful working environment during the excellent workshop \lq\lq From Scattering Amplitudes to the Conformal Bootstrap." M.~B. also gratefully acknowledges support from the Simons Center for Geometry and Physics, Stony Brook University during the interesting \lq\lq Simons Summer Workshop 2015." Our research was partially supported by the U.S. Department of Energy under grants DOE-SC0010008, DOE-ARRA-SC0003883, and DOE-DE-SC0007897. M.~B. is also partially supported by the U.S. Department of Energy under grant DE-SC0009924. T.~N. is also partially supported by the Yukawa Memorial Foundation.
}

\newpage

\begin{appendices}

\section{The Higgs branch character of the $(A_1,D_{2n})$ theory}
\label{app:Higgs-A1Dn}

In this appendix, we show that \eqref{eq:HL-AD} is identical to the character of the Higgs branch chiral ring of the $(A_1,D_{2n})$ theory. First of all, it is straightforward to rewrite \eqref{eq:HL-AD} as
\begin{align}\label{eq:HL-ADii}
\mathcal{I}_{(A_1,D_{2n})}(0,t;x,y) &=  \frac{1+t-t^{\frac{n}{2}+1}x(y+y^{-1})}{(1-t)(1-ty^2)(1-ty^{-2})(1-t^{\frac{n}{2}}xy)(1-t^{\frac{n}{2}}xy^{-1})}
\nonumber\\
& \qquad + \frac{1+t-t^{\frac{n}{2}+1}x^{-1}(y+y^{-1})}{(1-t)(1-ty^2)(1-ty^{-2})(1-t^{\frac{n}{2}}x^{-1}y)(1-t^{\frac{n}{2}}x^{-1}y^{-1})}
\nonumber\\
& \qquad -\frac{1+t}{(1-t)(1-ty^2)(1-ty^{-2})}~.
\end{align}
Now, recall that the Higgs branch chiral ring, $\mathcal{H}$, is generated by $M_i{}^j,\, L_i$ and $\tilde{L}^i$ subject to \eqref{eq:Higgs}.
Defining $\CH_1$ to be the sub-ring generated by $M_i^{\ j}$ and $L_i$ subject to the first two relations in \eqref{eq:Higgs} and $\CH_2$ to be the sub-ring generated by $M_i^{\ j}$ and $\tilde L^i$ subject to the first and third relations in \eqref{eq:Higgs}, we see that the last relation in \eqref{eq:Higgs} implies that $\CH=\CH_1\cup\CH_2$. 
As a result, the character of $\CH$ is just the sum of the characters of the $\CH_i$ minus the character of $\CH_1\cap\CH_2$, i.e., $\text{char}(\mathcal{H}) = \text{char}(\mathcal{H}_1) + \text{char}(\mathcal{H}_2) - \text{char}(\mathcal{H}_1\cap \mathcal{H}_2)$.

It is simple to check that the first term in \eqref{eq:HL-ADii} is the character of $\CH_1$. Indeed, we find
\begin{eqnarray}\label{H1char}
{\rm char}(\CH_1) &=&
{1-t^2-t^{{n\over2}+1}x(y+y^{-1})+t^{{n\over2}+2}x(y+y^{-1})\over(1-t)^2(1-ty^2)(1-ty^{-2})(1-t^{n\over2}xy)(1-t^{n\over2}xy^{-1})}\nonumber\\ &=&\frac{1+t-t^{\frac{n}{2}+1}x(y+y^{-1})}{(1-t)(1-ty^2)(1-ty^{-2})(1-t^{\frac{n}{2}}xy)(1-t^{\frac{n}{2}}xy^{-1})}~,
\end{eqnarray}
where the first equality can be justified as follows: the first term in the numerator of the RHS gives the character without relations, while the second and third terms impose the first and second relations in \eqref{eq:Higgs}. The fourth term in this numerator is to compensate for over-subtraction since both constraints set to null operators like $(\epsilon^{ik}\epsilon_{j\ell}M_i{}^jM_k{}^\ell) L_m = 2\epsilon_{j\ell}(\epsilon^{ik}M_i{}^jL_k)M_m{}^\ell$.

Note that the character of ${\rm char}(\CH_2)$ can be derived in a similar fashion and takes the form of \eqref{H1char} but with $x\to x^{-1}$. Finally, the character of $\CH_1\cap\CH_2$ is
\begin{align}
{\rm char}(\CH_1\cap\CH_2)=\frac{(1-t^2)}{(1-t)^2(1-ty^2)(1-ty^{-2})} = \frac{(1+t)}{(1-t)(1-ty^2)(1-ty^{-2})}~,
\end{align}
since it is generated by the $M_i^{\ j}$ subject to the first relation in \eqref{eq:Higgs}. Putting these derivations together, we find that the character of the Higgs branch chiral ring (i.e., the Hilbert series), $\text{char}(\mathcal{H})$, agrees with the HL index, $\mathcal{I}_{(A_1,D_{2n})}(0,t;x,y)$.

\section{Some chiral algebra matrix elements}
\label{app:MatrixElt}
Recall that in subsection \ref{sec:HRk}, we argued that there should not be any constraints of the form \eqref{ChirAlgAnalog} in the $(A_1, D_{2n})$ chiral algebras with $n>2$. To check this statement, we compute the following matrix elements in the singlet channel
\begin{eqnarray}\label{genericD2n}
\langle TJ|TJ\rangle&=&k_1\left(4+{c\over2}\right)=6-3n~,\nonumber
\\ \langle\partial J\cdot J|\partial J\cdot J\rangle&=&8k_1^2=8~,\nonumber\\ \langle\partial^2J|\partial^2J\rangle&=&12k_1=12~,\nonumber\\ \delta_{IJ}\delta_{KL}\langle J^I\partial J^J|J^K\partial J^L\rangle&=&6k(2+k)={6(1-2n)\over n^2}~, \\ \langle TJ|\partial^2J\rangle&=&6k_1=6~,\nonumber\\ \delta_{IJ}\langle\partial^2J|J^I\partial J^J\rangle&=&\langle TJ|\partial J\cdot J\rangle=\delta_{IJ}\langle\partial J\cdot J|J^I\partial J^J\rangle=\langle\partial J\cdot J|\partial^2J\rangle=\delta_{IJ}\langle TJ|J^I\partial J^J\rangle=0~,\nonumber
\end{eqnarray}
where we have set the $U(1)$ level to unity without loss of generality, and we have used $k=-2+{1\over n}$ and $c=-2(3n-2)$ \rcite{Buican:2015ina, Cordova:2015nma}. It is straightforward to see that the corresponding determinant is non-vanishing for all $n>2$. Similarly, in the adjoint channel, we have
\begin{eqnarray}\label{genericD2ntriplet}
\langle TJ^I|TJ^J\rangle&=&k\left(4+{c\over2}\right)\delta^{IJ}=\left(6n-15+{6\over n}\right)\delta^{IJ}~,\nonumber\\
\langle J\partial J^I|J\partial J^J\rangle&=&2k_1k\delta^{IJ}=2\left(-2+{1\over n}\right)\delta^{IJ}~,\nonumber\\ \langle \partial J\cdot J^I|\partial J\cdot J^J\rangle&=&4k_1k\delta^{IJ}=4\left(-2+{1\over n}\right)\delta^{IJ}~,\nonumber\\ \langle \left(J\partial J\right)^I|\left(J\partial J\right)^J\rangle&=& 8k\left(k+3\right)\delta^{IJ}=-8{(n+1)(2n-1)\over n^2}\delta^{IJ}~,\nonumber\\ \langle \partial^2 J^I|\partial^2J^J\rangle&=&12k\delta^{IJ}=12\left(-2+{1\over n}\right)\delta^{IJ}~,\nonumber \\ \langle (J\partial J)^I|TJ^J\rangle&=&-4k\delta^{IJ}=-4\left(-2+{1\over n}\right)\delta^{IJ}~,\\ \langle TJ^I|\partial^2 J^J\rangle&=&6k\delta^{IJ}=6\left(-2+{1\over n}\right)\delta^{IJ}~,\nonumber\\ \langle J\partial J^I|\partial J\cdot J^J \rangle&=&2k_1k\delta^{IJ}=2\left(-2+{1\over n}\right)\delta^{IJ}\nonumber~,\nonumber\\\langle(J\partial J)^I|\partial^2J^J\rangle&=&-16k\delta^{IJ}=-16\left(-2+{1\over n}\right)\delta^{IJ}~, \nonumber\\ \langle \partial J\cdot J^I|\partial^2J^J\rangle&=&\langle TJ^I|J\partial J^J\rangle=\langle TJ^I|\partial J\cdot J^J\rangle=\langle J\partial J^I|\partial^2 J^J\rangle=\langle \partial J\cdot J^I|\left(J\partial J\right)^J\rangle\nonumber~,\nonumber\\ &=& \langle J\partial J^I|\left(J\partial J\right)^J\rangle=0~.
\end{eqnarray}
It is again straightforward to check that the matrix of the above inner products has non-vanishing determinant. These results therefore serve as a useful consistency condition for our conjectures.

\section{Explicit check of our monopole inequality}
\label{app:Mineq}
In this appendix, we verify that the inequality \eqref{MonopoleEqn} holds in our theories. For ease of reference, we reproduce it below
\eqn{
\tilde E(\CO)>(I_3^R+\sum_a c^aH_a^{\CC})(\CO)=\sum_a c^aH_a^{\CC}(\CO)~,
}[MonopoleEqnii]
where $\CO$ is a monopole primary of the $S^1$ reduction under consideration. To prove that \eqref{MonopoleEqnii} holds, it is easier to check the following in the mirror theory
\eqn{
\tilde E(\tilde\CO)>\sum_a c^a\tilde H_a(\tilde\CO)~,
}[Monopolemirror]
where $\tilde\CO$ is the (matter) primary dual to $\CO$, and $\tilde H_a$ is the flavor symmetry dual to $H_a^{\CC}$.

First consider the $(A_1 A_{2n-3})$ theory. Recall from \rcite{Xie:2012hs} that the three-dimensional mirror of this theory can be reached by an RG flow from $\CN=4$ SQED with $N_f=n-1$. We denote the fundamental fields $X_I$ and the anti-fundamental $\CN=4$ partners $Y^I$ (with $I=1,\cdots, n-1$). To demonstrate \eqref{Monopolemirror}, it suffices to show it holds for all operators built from the squarks.

To prove this latter statement, recall that $c^a={1\over n\sqrt{2}}(-1)^{n+a}\sqrt{a(a+1)}$  \rcite{Buican:2015hsa}.
Therefore, we have
\begin{eqnarray}\label{valueii}
\sum_a c^a\tilde H_a(X_I)&=&{1\over n\sqrt{2}}\sum_a(-1)^{n+a}\sqrt{a(a+1)}\cdot \nu_{Ia}={1\over 2n}\Big((-1)^{n+I}\cdot(I-1)+\sum_{a=I}^{n-2}(-1)^{n+a}\Big)\nonumber\\&<&{1\over2}~,
\end{eqnarray}
where the $\nu_I$ are the weights for the fundamental representation of the $SU(n-1)$ flavor symmetry. Note that the same inequality holds for $X^{\dagger I}$, $Y^I$, and $Y^{\dagger}_I$. If $n+I$ is even, then, we have that the LHS of \eqref{valueii} is ${I\over2n}\le{n-2\over2n}<{1\over2}$. On the other hand, if $n+I$ is odd, we have that the LHS is ${-I+1\over2n}$ with $\left|{-I+1\over2n}\right|\le{n-3\over2n}<{1\over2}$. As a result, we see that all gauge-invariant matter operators built from the squark superfields satisfy \eqref{Monopolemirror}.\foot{The only way to get a gauge-invariant chiral matter operator to approach the bound in \eqref{Monopolemirror} is to take the limit $n\to\infty$ (however, the bound is not saturated for any finite $n$).}

Next let us consider the $(A_1, D_{2n})$ theory. Its mirror theory can be reached by an RG flow from a $U(1)^2$ quiver gauge theory with $X_I$ (and partners $Y^I$) charged under both $U(1)$ factors as well as $A$ (and partners $B$) and $\hat A$ (and partners $\hat B$) charged under different $U(1)$ factors \rcite{Xie:2012hs}. The $X_I$ and $Y^I$ are charged under an $SU(n-1)$ flavor symmetry as in the case of the $(A_1, A_{2n-3})$ theory, while the remaining fields are not. Therefore, any chiral matter operators we can build in this theory must also satisfy \eqref{Monopolemirror}.

\section{Formulas with $q$-binomial coefficients}
\label{app:q-series}

Here, we derive the expressions \eqref{eq:x=t-to-half} and \eqref{eq:sign-A1Dn}. To that end, we first recall the following proposition. Suppose that $a(u) =\sum_{\lambda=0}^\infty a_{\lambda} u^\lambda/(q;q)_\lambda$ and $b(u)=\sum_{\lambda=0}^\infty b_\lambda u^\lambda/(q;q)_\lambda$ are two absolutely convergent series. Then $a(u)b(u)$ can be expanded as 
\begin{align}
a(u)b(u) = \sum_{\lambda=0}^\infty \left(\sum_{m=0}^\lambda \frac{a_m b_{\lambda-m}}{(q;q)_m(q;q)_{\lambda-m}}\right)u^\lambda~.
\label{eq:ab}
\end{align}

Now, suppose that $a_\lambda = (t;q)_\lambda t^\lambda$ and $b_\lambda=(t;q)_\lambda$. For $|q|<1,\,|t|<1$ and $|u|<1$, it follows that $a(u) = (t^2u;q)_\infty/(tu;q)_\infty$ and $b(u) = (tu;q)_\infty/(u;q)_\infty$, thanks to the infinite $q$-binomial theorem: $\sum_{\lambda=0}^\infty \frac{(a;q)_\lambda}{(q;q)_\lambda}z^\lambda = \frac{(az;q)_\infty}{(z;q)_\infty}$. Therefore we have
\begin{align}
a(u)b(u) = \frac{(t^2u;q)_\infty}{(u;q)_\infty} = \sum_{\lambda=0}^\infty \frac{(t^2;q)_\lambda}{(q;q)_\lambda}u^\lambda~.
\end{align}
It follows from this result and \eqref{eq:ab} that
\begin{align}
 \sum_{m=0}^\lambda \frac{(t;q)_m(t;q)_{\lambda-m}}{(q;q)_m(q;q)_{\lambda-m}} t^{m} = \frac{(t^2;q)_\lambda}{(q;q)_\lambda}~,
\end{align}
which immediately implies \eqref{eq:x=t-to-half}.

Next, let us turn to the derivation of \eqref{eq:sign-A1Dn}. We now set $a_\lambda=(t;q)_\lambda y^{2\lambda}$ and $b_\lambda = (t;q)_\lambda$. For $|q|<1,\,|t|<1,\,|y|=1$ and $|u|<1$, we obtain
\begin{align}
a(u)b(u) = \frac{(ty^{2}u;q)_\infty (tu;q)_\infty}{(u;q)_\infty(y^{2}u;q)_\infty} = \left(\sum_{\lambda=0}^\infty \frac{(ty^{2};q)_\lambda}{(q;q)_\lambda}u^\lambda\right)\left(\sum_{\lambda=0}^\infty\frac{(ty^{-2};q)_\lambda}{(q;q)_\lambda}y^{2\lambda}u^\lambda\right)~.
\end{align}
This result and \eqref{eq:ab} implies that
\begin{align}
\sum_{m=0}^\lambda\frac{(t;q)_m(t;q)_{\lambda-m}}{(q;q)_{m}(q;q)_{\lambda-m}}y^{2m} = \sum_{m=0}^\lambda \frac{(ty^{-2};q)_m(ty^{2};q)_{\lambda-m}}{(q;q)_m(q;q)_{\lambda-m}}y^{2\lambda-2m}~,
\end{align}
which gives us the following alternative expression for the Macdonald polynomial:
\begin{align}
P_{\lambda}(q,t;y)=N_\lambda(q,t)\sum_{m=0}^\lambda \frac{(ty^{-2};q)_m(ty^2;q)_{\lambda-m}}{(q;q)_m(q;q)_{\lambda-m}}y^{\lambda-2m}~.
\label{eq:q-binomial}
\end{align}
Using this expression in \eqref{eq:first-A1Dn}, we obtain
\begin{align}
\mathcal{I}_{(A_1,D_{2n})}(q,t;x,y) 
=& \frac{1}{(t;q)_\infty}\sum_{\lambda=0}^\infty\frac{t^{\frac{n\lambda}{2}}}{(q;q)_\lambda} \sum_{m_1,m_2=0}^\lambda\left[\!\!
\begin{array}{c}
\lambda\\
m_1\\
\end{array}
\!\!\right]_q \left[\!\!
\begin{array}{c}
\lambda\\
m_2\\
\end{array}
\!\!\right]_q
\frac{(1-q^\lambda t)}{(q^{m_1}t;q)_\infty(q^{\lambda-m_1}t;q)_\infty}
\nonumber\\
&\quad \times \frac{(q^\lambda t^2;q)_\infty}{(q^{m_2}ty^{-2};q)_\infty(q^{\lambda-m_2}ty^2;q)_\infty}q^{nm_1(\lambda-m_1)}x^{2m_1-\lambda}y^{\lambda-2m_2}~,
\end{align}
where $\scriptsize{\left[\!\!
\begin{array}{c}
\lambda \\
m\\
\end{array}
\!\!\right]_q} \equiv (q;q)_\lambda/\{(q;q)_m(q;q)_{\lambda-m}\}$. Here, the infinite $q$-binomial theorem implies
\begin{align}
\frac{(q^\lambda t^2;q)_\infty}{(q^{m_2}ty^{-2};q)_\infty(q^{\lambda-m_2}ty^2;q)_\infty} 
= \sum_{k=0}^\infty \frac{(q^{\lambda-m_2}ty^2)^k}{(q;q)_k(q^{m_2+k}ty^{-2};q)_\infty}~.
\end{align}
Combining the above two equations, we finally obtain the desired expression \eqref{eq:sign-A1Dn}.

\section{The pole structure of the $(A_1,D_{2n})$ index}
\label{app:poles-A1Dn}

In this appendix, we will derive the expression \eqref{eq:resum2} for the $(A_1,D_{2n})$ index.
Let us start with the expression in \eqref{eq:first-A1Dn}. By exchanging the order of the summations, it can be rewritten as
\begin{align}
\mathcal{I}_{(A_1,D_{2n})}(q,t;x,y) 
=& \frac{(t^2;q)_\infty}{(t;q)_\infty^3 (ty^2;q)_\infty(t/y^2;q)_\infty}\sum_{m_1=0}^\infty\sum_{m_2=0}^\infty \frac{(t;q)_{m_1}(t;q)_{m_2}}{(q;q)_{m_1}(q;q)_{m_2}} q^{-nm_1^2}x^{2m_1}y^{-2m_2}
\nonumber\\
&\quad \times \sum_{\lambda = \text{max}(m_1,m_2)}^\infty \!\!\!\!\! \frac{(1-tq^\lambda)(q;q)_\lambda}{(t^2;q)_\lambda}\frac{(t;q)_{\lambda-m_1}(t;q)_{\lambda-m_2}}{(q;q)_{\lambda-m_1}(q;q)_{\lambda-m_2}}\left(q^{nm_1}t^{\frac{n}{2}} x^{-1} y\right)^\lambda~.
\label{eq:second-A1Dn}
\end{align}
Note here that, in the case of $m_1 >  m_2$, the sum over $\lambda$ is written as
\begin{align}
&\frac{(q;q)_{m_1}(t;q)_{m_1-m_2}}{(t^2;q)_{m_1}(q;q)_{m_1-m_2}}(1-tq^{m_1})\left(q^{nm_1}t^{\frac{n}{2}}x^{-1}y\right)^{m_1}
\nonumber\\
&\qquad \qquad \times {}_4\varphi_3\left(
\begin{array}{cccc}
t &  tq^{m_1+1} & q^{m_1+1} & tq^{m_1-m_2}\\
&  tq^{m_1} & t^2q^{m_1} & q^{m_1-m_2+1}\\
\end{array}
;\; q^{nm_1}t^{\frac{n}{2}}x^{-1}y\right)~,
\label{eq:m1-m2}
\end{align}
where ${}_4\varphi_3$ is the basic hypergeometric series given by
\begin{align}
{}_4\varphi_3\left(
\begin{array}{cccc}
a_0 & a_1 & a_2 & a_3\\
 & b_1 & b_2 & b_3\\
\end{array}
;\; z\right) \equiv \sum_{\lambda=0}^\infty \frac{\prod_{k=0}^{3}(a_k;q)_\lambda}{(q;q)_\lambda\prod_{k=1}^{3}(b_k;q)_\lambda}z^\lambda~.
\label{eq:basic-hyper}
\end{align}
On the other hand, in the case of $m_1\leq m_2$, the sum over $\lambda$ becomes
\begin{align}
&\frac{(q;q)_{m_2}(t;q)_{m_2-m_1}}{(t^2;q)_{m_2}(q;q)_{m_2-m_1}}(1-tq^{m_2})(q^{nm_1}t^{\frac{n}{2}}x^{-1}y)^{m_2}
\nonumber\\
&\qquad \qquad \times {}_4\varphi_3\left(
\begin{array}{cccc}
t & tq^{m_2+1} & q^{m_2+1} & tq^{m_2-m_1}\\
  & tq^{m_2} & t^2q^{m_2} & q^{m_2-m_1+1}\\
\end{array}
;\; q^{nm_1}t^{\frac{n}{2}}x^{-1}y\right)~.
\label{eq:m2-m1}
\end{align}
The expressions \eqref{eq:m1-m2} and \eqref{eq:m2-m1} implies that \eqref{eq:second-A1Dn} is rewritten as \eqref{eq:resum2}.

\end{appendices}

\newpage
\bibliography{chetdocbib}

\begin{thebibliography}{10}
\ifx\href\asklfhas\newcommand{\href}[2]{#2}\fi
\ifx\arxivref\asklfhas\newcommand{\arxivref}[2]{\href{http://arxiv.org/abs/#1}{#2}}\fi
\ifx\doiref\asklfhas\newcommand{\doiref}[2]{\href{http://dx.doi.org/#1}{#2}}\fi
\parskip 0pt
\normalsize

\bibitem{Argyres:1995jj}
P.C. Argyres \& M.R. Douglas,
\textit{``{New phenomena in SU(3) supersymmetric gauge theory}''},
\doiref{10.1016/0550-3213(95)00281-V}{Nucl.Phys. \textbf{B448}, 93 (1995)},
\normalsize{\texttt{\arxivref{hep-th/9505062}{hep-th/9505062}}}.

\bibitem{Argyres:1995xn}
P.C. Argyres, M.R. Plesser, N.~Seiberg \& E.~Witten,
\textit{``{New N=2 superconformal field theories in four-dimensions}''},
\doiref{10.1016/0550-3213(95)00671-0}{Nucl.Phys. \textbf{B461}, 71 (1996)},
\normalsize{\texttt{\arxivref{hep-th/9511154}{hep-th/9511154}}}.

\bibitem{Eguchi:1996vu}
T.~Eguchi, K.~Hori, K.~Ito \& S.K. Yang,
\textit{``{Study of N=2 superconformal field theories in four-dimensions}''},
\doiref{10.1016/0550-3213(96)00188-5}{Nucl.Phys. \textbf{B471}, 430 (1996)},
\normalsize{\texttt{\arxivref{hep-th/9603002}{hep-th/9603002}}}.

\bibitem{Gaiotto:2010jf}
D.~Gaiotto, N.~Seiberg \& Y.~Tachikawa,
\textit{``{Comments on scaling limits of 4d N=2 theories}''},
\doiref{10.1007/JHEP01(2011)078}{JHEP \textbf{1101}, 078 (2011)},
\normalsize{\texttt{\arxivref{1011.4568}{arXiv:1011.4568}}}.

\bibitem{Buican:2015ina}
M.~Buican \& T.~Nishinaka,
\textit{``{On the Superconformal Index of Argyres-Douglas Theories}''},
\normalsize{\texttt{\arxivref{1505.05884}{arXiv:1505.05884}}}.

\bibitem{Buican:2015hsa}
M.~Buican \& T.~Nishinaka,
\textit{``{Argyres-Douglas Theories, $S^1$ Reductions, and Topological
  Symmetries}''},
\normalsize{\texttt{\arxivref{1505.06205}{arXiv:1505.06205}}}.

\bibitem{Cordova:2015nma}
C.~Cordova \& S.H. Shao,
\textit{``{Schur Indices, BPS Particles, and Argyres-Douglas Theories}''},
\normalsize{\texttt{\arxivref{1506.00265}{arXiv:1506.00265}}}.

\bibitem{Bonelli:2011aa}
G.~Bonelli, K.~Maruyoshi \& A.~Tanzini,
\textit{``{Wild Quiver Gauge Theories}''},
\doiref{10.1007/JHEP02(2012)031}{JHEP \textbf{1202}, 031 (2012)},
\normalsize{\texttt{\arxivref{1112.1691}{arXiv:1112.1691}}}.

\bibitem{Xie:2012hs}
D.~Xie,
\textit{``{General Argyres-Douglas Theory}''},
\doiref{10.1007/JHEP01(2013)100}{JHEP \textbf{1301}, 100 (2013)},
\normalsize{\texttt{\arxivref{1204.2270}{arXiv:1204.2270}}}.

\bibitem{Gadde:2011ik}
A.~Gadde, L.~Rastelli, S.S. Razamat \& W.~Yan,
\textit{``{The 4d Superconformal Index from q-deformed 2d Yang-Mills}''},
\doiref{10.1103/PhysRevLett.106.241602}{Phys.Rev.Lett. \textbf{106}, 241602
  (2011)},
\normalsize{\texttt{\arxivref{1104.3850}{arXiv:1104.3850}}}.

\bibitem{Wang:2015mra}
Y.~Wang \& D.~Xie,
\textit{``{Classification of Argyres-Douglas theories from M5 branes}''},
\normalsize{\texttt{\arxivref{1509.00847}{arXiv:1509.00847}}}.

\bibitem{Gadde:2011uv}
A.~Gadde, L.~Rastelli, S.S. Razamat \& W.~Yan,
\textit{``{Gauge Theories and Macdonald Polynomials}''},
\doiref{10.1007/s00220-012-1607-8}{Commun.Math.Phys. \textbf{319}, 147 (2013)},
\normalsize{\texttt{\arxivref{1110.3740}{arXiv:1110.3740}}}.

\bibitem{Buican:2014qla}
M.~Buican, T.~Nishinaka \& C.~Papageorgakis,
\textit{``{Constraints on chiral operators in $ \mathcal{N}=2 $ SCFTs}''},
\doiref{10.1007/JHEP12(2014)095}{JHEP \textbf{1412}, 095 (2014)},
\normalsize{\texttt{\arxivref{1407.2835}{arXiv:1407.2835}}}.

\bibitem{Dolan:2002zh}
F.~Dolan \& H.~Osborn,
\textit{``{On short and semi-short representations for four-dimensional
  superconformal symmetry}''},
\doiref{10.1016/S0003-4916(03)00074-5}{Annals~Phys. \textbf{307}, 41 (2003)},
\normalsize{\texttt{\arxivref{hep-th/0209056}{hep-th/0209056}}}.

\bibitem{Dobrev:1985qv}
V.~Dobrev \& V.~Petkova,
\textit{``{All Positive Energy Unitary Irreducible Representations of Extended
  Conformal Supersymmetry}''},
\doiref{10.1016/0370-2693(85)91073-1}{Phys.Lett. \textbf{B162}, 127 (1985)}.

\bibitem{toappear}
M.~Buican \& T.~Nishinaka,
\textit{``{On Operator Constraints in Argyres-Douglas Theories}''},
\normalsize{\texttt{\arxivref{15XX.XXXXX}{arXiv:15XX.XXXXX}}}.

\bibitem{DelZotto:2014kka}
M.~Del~Zotto \& A.~Hanany,
\textit{``{Complete Graphs, Hilbert Series, and the Higgs branch of the 4d
  $\mathcal{N} =$ 2 $(A_n,A_m)$ SCFTs}''},
\doiref{10.1016/j.nuclphysb.2015.03.017}{Nucl.~Phys. \textbf{B894}, 439
  (2015)},
\normalsize{\texttt{\arxivref{1403.6523}{arXiv:1403.6523}}}.

\bibitem{Beem:2013sza}
C.~Beem, M.~Lemos, P.~Liendo, W.~Peelaers, L.~Rastelli et~al.,
\textit{``{Infinite Chiral Symmetry in Four Dimensions}''},
\normalsize{\texttt{\arxivref{1312.5344}{arXiv:1312.5344}}}.

\bibitem{Buican:2014hfa}
M.~Buican, S.~Giacomelli, T.~Nishinaka \& C.~Papageorgakis,
\textit{``{Argyres-Douglas Theories and S-Duality}''},
\doiref{10.1007/JHEP02(2015)185}{JHEP \textbf{1502}, 185 (2015)},
\normalsize{\texttt{\arxivref{1411.6026}{arXiv:1411.6026}}}.

\bibitem{DelZotto:2015rca}
M.~Del~Zotto, C.~Vafa \& D.~Xie,
\textit{``{Geometric Engineering, Mirror Symmetry and 6d (1,0) $\to$ 4d,
  N=2}''},
\normalsize{\texttt{\arxivref{1504.08348}{arXiv:1504.08348}}}.

\bibitem{Cecotti:2015hca}
S.~Cecotti \& M.~Del~Zotto,
\textit{``{Higher S-dualities and Shephard-Todd groups}''},
\normalsize{\texttt{\arxivref{1507.01799}{arXiv:1507.01799}}}.

\bibitem{Cecotti:2012va}
S.~Cecotti,
\textit{``{Categorical Tinkertoys for N=2 Gauge Theories}''},
\doiref{10.1142/S0217751X13300068}{Int.~J.~Mod.~Phys. \textbf{A28}, 1330006
  (2013)},
\normalsize{\texttt{\arxivref{1203.6734}{arXiv:1203.6734}}}.

\bibitem{Gaiotto:2012sf}
D.~Gaiotto \& J.~Teschner,
\textit{``{Irregular singularities in Liouville theory and Argyres-Douglas type
  gauge theories, I}''},
\doiref{10.1007/JHEP12(2012)050}{JHEP \textbf{1212}, 050 (2012)},
\normalsize{\texttt{\arxivref{1203.1052}{arXiv:1203.1052}}}.

\bibitem{Kanno:2013vi}
H.~Kanno, K.~Maruyoshi, S.~Shiba \& M.~Taki,
\textit{``{$W_3$ irregular states and isolated $N=2$ superconformal field
  theories}''},
\doiref{10.1007/JHEP03(2013)147}{JHEP \textbf{1303}, 147 (2013)},
\normalsize{\texttt{\arxivref{1301.0721}{arXiv:1301.0721}}}.

\bibitem{Argyres:2012fu}
P.C. Argyres, K.~Maruyoshi \& Y.~Tachikawa,
\textit{``{Quantum Higgs branches of isolated N=2 superconformal field
  theories}''},
\doiref{10.1007/JHEP10(2012)054}{JHEP \textbf{1210}, 054 (2012)},
\normalsize{\texttt{\arxivref{1206.4700}{arXiv:1206.4700}}}.

\bibitem{Gaiotto:2012xa}
D.~Gaiotto, L.~Rastelli \& S.S. Razamat,
\textit{``{Bootstrapping the superconformal index with surface defects}''},
\doiref{10.1007/JHEP01(2013)022}{JHEP \textbf{1301}, 022 (2013)},
\normalsize{\texttt{\arxivref{1207.3577}{arXiv:1207.3577}}}.

\bibitem{Bowman}
D.~Bowman,
\textit{``{A general Heine transformation and symmetric polynomials of
  Rogers}''},
\doiref{10.1007/978-1-4612-4086-0_9}{Analytic~Number~Theory;~Progress~in~Mathematics
  \textbf{138}, 163 (1996)}.

\bibitem{Benini:2011nc}
F.~Benini, T.~Nishioka \& M.~Yamazaki,
\textit{``{4d Index to 3d Index and 2d TQFT}''},
\doiref{10.1103/PhysRevD.86.065015}{Phys.~Rev. \textbf{D86}, 065015 (2012)},
\normalsize{\texttt{\arxivref{1109.0283}{arXiv:1109.0283}}}.

\bibitem{Razamat:2014pta}
S.S. Razamat \& B.~Willett,
\textit{``{Down the rabbit hole with theories of class $ \mathcal{S} $}''},
\doiref{10.1007/JHEP10(2014)099}{JHEP \textbf{1410}, 99 (2014)},
\normalsize{\texttt{\arxivref{1403.6107}{arXiv:1403.6107}}}.

\bibitem{Liendo:2015ofa}
P.~Liendo, I.~Ramirez \& J.~Seo,
\textit{``{Stress-tensor OPE in N=2 Superconformal Theories}''},
\normalsize{\texttt{\arxivref{1509.00033}{arXiv:1509.00033}}}.

\bibitem{Zamolodchikov:1986gt}
A.B. Zamolodchikov,
\textit{``{Irreversibility of the Flux of the Renormalization Group in a 2D
  Field Theory}''},
JETP~Lett. \textbf{43}, 730 (1986),
[Pisma Zh. Eksp. Teor. Fiz.43,565(1986)].

\bibitem{Komargodski:2011vj}
Z.~Komargodski \& A.~Schwimmer,
\textit{``{On Renormalization Group Flows in Four Dimensions}''},
\doiref{10.1007/JHEP12(2011)099}{JHEP \textbf{1112}, 099 (2011)},
\normalsize{\texttt{\arxivref{1107.3987}{arXiv:1107.3987}}}.

\bibitem{Komargodski:2011xv}
Z.~Komargodski,
\textit{``{The Constraints of Conformal Symmetry on RG Flows}''},
\doiref{10.1007/JHEP07(2012)069}{JHEP \textbf{1207}, 069 (2012)},
\normalsize{\texttt{\arxivref{1112.4538}{arXiv:1112.4538}}}.

\bibitem{Intriligator:2005if}
K.A. Intriligator,
\textit{``{IR free or interacting? A Proposed diagnostic}''},
\doiref{10.1016/j.nuclphysb.2005.10.005}{Nucl.~Phys. \textbf{B730}, 239
  (2005)},
\normalsize{\texttt{\arxivref{hep-th/0509085}{hep-th/0509085}}}.

\bibitem{Buican:2011ty}
M.~Buican,
\textit{``{A Conjectured Bound on Accidental Symmetries}''},
\doiref{10.1103/PhysRevD.85.025020}{Phys.~Rev. \textbf{D85}, 025020 (2012)},
\normalsize{\texttt{\arxivref{1109.3279}{arXiv:1109.3279}}}.

\bibitem{Buican:2012ec}
M.~Buican,
\textit{``{Non-Perturbative Constraints on Light Sparticles from Properties of
  the RG Flow}''},
\doiref{10.1007/JHEP10(2014)026}{JHEP \textbf{1410}, 26 (2014)},
\normalsize{\texttt{\arxivref{1206.3033}{arXiv:1206.3033}}}.

\end{thebibliography}

\end{document}